\documentclass[%
 reprint,
 superscriptaddress,
 amsmath,amssymb,
 aps,
 pra
]{revtex4-1}

\usepackage{braket}
\usepackage{graphicx}
\usepackage{dcolumn}
\usepackage{bm}
\usepackage{siunitx}
\usepackage{xcolor} 
\usepackage{tabularx}
\usepackage{dsfont}					
\usepackage{mathrsfs}				
\usepackage{float}
\usepackage{hyperref}

\newcommand{\comment}[1]{}			

\newcommand{\be}[0]{\begin{equation}}	
\newcommand{\ee}[0]{\end{equation}}

\newcommand{\sgn}{\mbox{sgn}}
\newcommand{\rmi}{{\rm i}}

\begin{document}


\title{Critical Theory for Breakdown of Photon Blockade}

\author{Jonathan B. Curtis}
 \email{jcurtis1@umd.edu}
\affiliation{Joint Quantum Institute, NIST/University of Maryland, College Park, Maryland 20742, USA}
 \author{Igor Boettcher}
 \affiliation{Joint Quantum Institute, NIST/University of Maryland, College Park, Maryland 20742, USA}
\author{Jeremy T. Young}
\affiliation{Joint Quantum Institute, NIST/University of Maryland, College Park, Maryland 20742, USA}
\author{Mohammad F. Maghrebi}
\affiliation{Department of Physics and Astronomy, Michigan State University, East Lansing, Michigan 48824, USA}
\author{Howard Carmichael}
\affiliation{The Dodd-Walls Centre for Photonic and Quantum Technologies, Department of Physics, University of Auckland, Private Bag 92019, Auckland, New Zealand}
\author{Alexey V. Gorshkov}
\affiliation{Joint Quantum Institute, NIST/University of Maryland, College Park, Maryland 20742, USA}
\affiliation{Joint Center for Quantum Information and Computer Science, NIST/University of Maryland, College Park, Maryland 20742, USA}
\author{Michael Foss-Feig}
\affiliation{United States Army Research Laboratory, Adelphi, Maryland 20783, USA}

\date{\today}

\begin{abstract}
Photon blockade is the result of the interplay between the quantized nature of light and strong optical nonlinearities, whereby strong photon-photon repulsion prevents a quantum optical system from absorbing multiple photons. 
We theoretically study a single atom coupled to the light field, described by the resonantly driven Jaynes--Cummings model, in which case the photon blockade breaks down in a second order phase transition at a critical drive strength.
We show that this transition is associated to the spontaneous breaking of an anti-unitary $\mathcal{PT}$-symmetry.
Within a semiclassical approximation we calculate the expectation values of observables in the steady state. 
We then move beyond the semiclassical approximation and approach the critical point from the disordered (blockaded) phase by reducing the Lindblad quantum master equation to a classical rate equation that we solve.
The width of the steady-state distribution in Fock space is found to diverge as we approach the critical point with a simple power-law, allowing us to calculate the critical scaling of steady state observables without invoking mean-field theory. 
We propose a simple physical toy model for biased diffusion in the space of occupation numbers, which captures the universal properties of the steady state.
We list several experimental platforms where this phenomenon may be observed.

\end{abstract}
\maketitle

\section{\label{sec:intro}Introduction}

Thermal equilibrium is an incredibly powerful and constraining property of a large class of quantum many-body systems.
Quantum systems departing from equilibrium often exhibit rich novel physics, including phenomena such as many-body localization~\cite{Nandkishore2015,Abanin2019,Kohlert2019}, many-body scars~\cite{Bernien2017,Turner2018}, time-crystalline order~\cite{Wilczek2012,Watanabe2015,Else2016,Else2017,Zhang2016a,Choi2017}, exotic Floquet order~\cite{Potirniche2017,Potter2017,Oka2009,Lindner2011,Kitagawa2010}, dynamical phase transitions~\cite{Heyl2018}, and superradiance~\cite{Baumann2010,Buchhold2013}.
Experimentally, there is a wide array of platforms available for realizing these non-equilibrium phenomena, including optical-tweezer arrays of Ryberg atoms~\cite{Bernien2017}, ultra-cold atoms~\cite{Kohlert2019} and molecules \cite{yan13} in optical lattices, Bose-Einstein condensates~\cite{Baumann2010}, trapped ions~\cite{Zhang2016a}, optical defects in diamond~\cite{Choi2017,Arcizet2011,Gieseler2019}, exciton-polariton condensates~\cite{Kasprzak2006}, semiconductor quantum dots~\cite{Faraon2008}, and interacting circuit or cavity photons~\cite{Birnbaum2005,Bishop2009,Lang2011,Raftery2014,Fink2017,Fitzpatrick2017,Vaneph2018,Snijders2018}.
The ability to precisely control and measure quantum systems out of equilibrium is also a key component in a number of emerging quantum technologies~\cite{Ge2019,Toyoda2013,Debnath2018,Majumdar2012,Faraon2008,Gieseler2019,Groblacher2009,Crespi2012,Rabl2009,Huillery2020}.

Deviations from the equilibrium ensemble are particularly common in the field of quantum optics~\cite{Drummond1980,Alsing1991,Alsing1992,Dutra1993,Dombi2015,Carmichael2015,Mendoza-Arenas2016,Maghrebi16,Mavrogordatos2016,Foss-Feig2017,Young2020,Marino2016a,Marino2016,Rota2019}.
Since the typical temperature of a black-body that emits in the optical frequency range (roughly \SI{500}{THz}) is of order \SI{4800}{K}, almost all experiments involving optical photons are conducted far from the equilibrium temperature scale~\footnote{This argument implicitly relies on the fact that photons have no chemical potential. Thus, far below their characteristic energy scale, they are simple non-existent.}.
Furthermore, processes of interest often involve strong (coherent) driving, e.g.~by lasers, in part to overcome the losses due to the fact that the system of interest is typically coupled to a highly incoherent environment. 
This combination of strong driving and incoherent loss processes often results in a quantum system which is far from thermal equilibrium.

Non-equilibrium effects are especially pronounced in dynamics of systems with strong interactions.
For instance, strong optical nonlinearities give rise to the photon blockade~\cite{Imamoglu97,Greentree2006,Angelakis2007,Birnbaum2005,Faraon2008,Shamailov2010,Lang2011,Majumdar2012,hoffman11}, whereby strong photon-photon repulsion inhibits the absorption of more than one photon, even in the presence of strong external driving.
In this case, the effective single-occupancy constraint becomes readily apparent in photon transport through the device.

Photon blockade can be understood qualitatively by considering a single-mode of an electromagnetic resonator with non-linear spectrum $E_n = \omega_0 n + \frac12 U n(n-1)$ as a function of photon number $n$ (i.e.~Hubbard term in the case of a lattice). 
To lowest order in perturbation theory, coherently driving the photon field at frequency $\omega$ will connect a Fock state with $n$ photons to a state with $n+1$ photons, resulting in a correction to the eigenstates with an energy denominator $\sim E_{n+1} -E_n - \omega $. 
For the Kerr-type of non-linearity described above, this can result in at most one divergent term, which occurs when the drive frequency is tuned to satisfy $\omega - \omega_0 - Un = 0 $. 
The net result of this effect is that for sufficiently strong non-linearities $U$, the driven resonator effectively becomes a two-level system, indicating a strong photon anti-bunching and hence a departure from the thermal ensemble with Poissonian photon number statistics. 

A convenient way of engineering a strong optical non-linearity is to couple the resonator photon to an atom, which acts as a strongly non-linear ``hard-core" boson.
If we describe the system using the basic Jaynes-Cummings model~\cite{Jaynes1963} (see Sec.~\ref{sec:model}), we find the celebrated $\sqrt{n}$-non-linearity, resulting in the excitation spectrum $E_n \sim \omega_0 n \pm g \sqrt{n}$, with $g$ the atom-photon coupling constant.
As first observed in Ref.~\cite{Carmichael2015}, this has important implications for the fate of the photon-blockade.
Re-tooling the argument from the previous section, we see that there are divergences in the perturbative expansion when the driving frequency satisfies $\omega - \omega_0 \pm g \left( \sqrt{n+1} - \sqrt{n}\right) =0$.
For $\omega = \omega_0$, the non-linearity vanishes for $n\to \infty$ as $n^{-1/2}$.
This implies that the driven oscillator can effectively tunnel off to a high-photon number state, for which the non-linearity is less important.

Remarkably, it appears that the breakdown of the photon-blockade proceeds via a continuous dissipative phase transition~\cite{Carmichael2015,Gutierrez-Jauregui2018}.
More specifically, this phase transition presents as a non-analyticity in the steady state of the Lindblad equation which governs the driven-dissipative dynamics of the system.
Since the system we consider is a single atom coupled to a single-mode of the electromagnetic field, it constitutes a system with no spatial extent. 
It is then not clear {\it a priori} what the relevant ``system-size" parameter is, and what the appropriate thermodynamic limit is.
Since the Fock space of the photon is formally infinite, we can see that non-analyticities may arise if the entirety of the Fock space is accessible. 
As we discuss in more detail later, this identifies the thermodynamic limit with the limit of vanishing dissipation~\cite{Paz,Casteels2017,Rota2019}. 
It is the purpose of this work to provide a comprehensive analysis of the nature of this critical point.
The main results can be found in Tab.~\ref{tab:main-results}, which presents the predicted critical scaling of various steady-state expectation values of observables as the critical point is approached.
The approach to the critical point is controlled by the dimensionless parameter $\epsilon = 2\mathcal{E}/g$, which measures the drive strength $\mathcal{E}$ relative to the strength of the atom-photon coupling constant $g$. 
Criticality occurs at $\epsilon^2 = 1$, with $\epsilon^2 < 1$ corresponding to the disordered phase.

\renewcommand{\arraystretch}{1.4}
\begin{table}[t]
    \begin{tabular}{| c | c | }
    \hline
    \ Observable \   &   Scaling  \\
    \hline
    $\langle \hat{a}\rangle $& 0 \\
    $\langle \hat{a}^\dagger \hat{a} \rangle $& \ $(1-\epsilon^2)^{-1}$ \ \\
    \hline 
$\langle (\hat{\sigma}_x, \hat{\sigma}_y, \hat{\sigma}_z) \rangle $ & $\ (-\epsilon,\ 0,\ 1-\epsilon^2)\ $ \\
    $ |\langle \hat{\sigma}_x \rangle|^2 + |\langle \hat{\sigma}_y \rangle|^2+ |\langle \hat{\sigma}_z \rangle|^2$ & $ 1 $ \\
    \hline
    \end{tabular}
    \caption{Scaling of observables as we approach the critical point. 
    The parameter which controls the distance to the critical point is $\epsilon = 2\mathcal{E}/g$, with $\mathcal{E}$ the coherent drive strength and $g$ the atom-photon coupling strength.
    The critical point is located at $\epsilon^2 =1 $, with $\epsilon^2 < 1$ the disordered phase.
    This applies in the thermodynamic limit, which in this case is obtained by taking the zero-dissipation limit of the steady-state ensemble (see Sec.~\ref{sec:rateeqn}). }
    \label{tab:main-results}
\end{table}
\renewcommand{\arraystretch}{1}

This work is structured as follows.
In Sec.~\ref{sec:model}, we discuss the model in more detail, focusing on the coherent portion of the evolution.
In Sec.~\ref{sec:mft}, we  present a semi-classical treatment of the problem and a brief description of the physical intuition behind the critical point. 
The main calculation is contained in Sec.~\ref{sec:rateeqn}, where we use a rate equation to solve for the steady-state behavior.
Finally, in Sec.~\ref{sec:steady-state},  we use the results of Sec.~\ref{sec:rateeqn} to compute the critical scaling of physical observables near the critical point.
We conclude in Sec.~\ref{sec:outlook} with a discussion of the implications of our calculation as well as some future directions of interest. More technical aspects of the analysis are presented in the appendices.

\section{\label{sec:model}Model}

\subsection{Hamiltonian and master equation}

The model we consider in this work is that of an atom coupled to a single electromagnetic mode of a cavity. 
The atom is modeled as a two-level system, with ground state $\ket{\downarrow}$ and excited state $\ket{\uparrow}$.
For simplicity we assume, as in Ref.~\cite{Carmichael2015}, that the bare atomic and cavity frequencies are tuned to resonance with each other. 
Generically, the critical theory describing the breakdown of photon blockade exhibits a first-order phase transition when the detuning of the cavity drive is tuned away from being on resonance with the atom and cavity frequencies.
Only when all three frequencies are resonant does the system exhibit a second order phase transition. 
We restrict our focus to the continuous phase transition and assume that all three of the bare frequencies are tuned to resonance at $\omega_0$. 
We model the atom-photon interaction by a simple dipole transition with coupling constant $g$ and assume the rotating-wave approximation to be applicable.

The intrinsic coherent dynamics is described by the celebrated Jaynes--Cummings model \cite{Jaynes1963,Shore1993} with Hamiltonian
\begin{equation}
\label{eqn:JC}
\hat{H}_0 = \omega_0( \hat{a}^\dagger \hat{a} + \hat{\sigma}_+\hat{\sigma}_{-} ) +g (\hat{a}^\dagger \hat{\sigma}_{-} + \hat{a} \hat{\sigma}_{+}).
\end{equation}
Herein, $\hat{a}$ is the photon annihilation operator and $\hat{\sigma}_{\pm} = \frac12 (\hat{\sigma}_x \pm i \hat{\sigma}_y)$ are the atomic transition operators written in terms of Pauli matrices. In order to model the external coherent driving, e.g.\ by a laser imposed onto the cavity, we add the term
\begin{equation}
    \label{eqn:drive}
    \hat{H}_{\textrm{d}}(t) = \mathcal{E}(\hat{a}e^{i\omega_0 t} + \hat{a}^\dagger e^{-i\omega_0 t})
\end{equation}
to the Hamiltonian. This corresponds to a monochromatic driving of the cavity photon field at resonance frequency $\omega_0$ with strength $\mathcal{E}>0$. We incorporate dissipation through single-photon loss from the cavity to a zero temperature reservoir at rate $2\kappa$. The dynamics of the density matrix is then given by the open-system Lindblad master equation
\begin{equation}
    \label{eqn:lindblad}
    \frac{\partial \hat{\rho}}{\partial t} = - i[ \hat{H}_0 + \hat{H}_{\textrm{d}}(t),\hat{\rho}] + \kappa\left( 2 \hat{a}\hat{\rho} \hat{a}^\dagger -\{\hat{a}^\dagger \hat{a} ,\hat{\rho}\} \right).
\end{equation}
We are particularly interested in the steady-state solutions of this equation.

Let us briefly review the diagonalization of the Jaynes--Cummings Hamiltonian $\hat{H}_0$.
Due to the rotating-wave approximation,  the ``polariton" number $\hat{N} = \hat{a}^\dagger \hat{a} + \hat{\sigma}_+\hat{\sigma}_{-}$ is conserved.
Thus, the eigenspectrum of $\hat{H}_0$ separates into a direct sum of decoupled two-level systems, consisting of entangled atom-matter excitations known as polaritons.
The energy eigenstates of $\hat{H}_0$ are labeled by a single signed quantum number $\nu = 0, \pm \sqrt{n}$ with $n = 1,2,...$ and read
\begin{equation}
    \label{eqn:JCeigen}
    |\nu\rangle_0 = \begin{cases}
    |\textrm{vac}\rangle\otimes\ket{\downarrow} & \nu =  0 \\
    \frac{1}{\sqrt{2}}\Bigl(|n\rangle\otimes\ket{\downarrow}\pm |n-1\rangle\otimes\ket{\uparrow}\Bigr) & \nu \neq 0 .\\
    \end{cases}
\end{equation}
These states have energies $E_0^{(0)} = 0$ and $ E_{\nu}^{(0)} = \omega_0 n \pm g \sqrt{n}$, respectively.
In the following, we omit the direct product symbol.

In order to understanding the steady state of Eq.\ \eqref{eqn:lindblad}, it is helpful to first focus on the coherent evolution generated by $\hat{H}_0+\hat{H}_{\rm d}(t)$, neglecting dissipation.
Due to the time-dependent drive $\hat{H}_{\textrm{d}}(t)$, neither energy nor polariton number is conserved.
However, we may still utilize the $U(1)$ symmetry generated by $\hat{N}$ to analyze the system.
By applying the unitary operator $\hat{U}_{\textrm{d}}(t) = \exp(-i\omega_0 t \hat{N})$, we can go to a frame co-rotating with the drive frequency.
The resulting rotating-frame Hamiltonian
\begin{equation}
\label{eqn:rotating-frame}
\hat{H}_{\textrm{rf}} = g (\hat{a}^\dagger \hat{\sigma}_{-} + \hat{a} \hat{\sigma}_{+} ) + \mathcal{E}(\hat{a}+ \hat{a}^\dagger)
\end{equation}
is time-independent and, because the drive is resonant, linear in the photon operators. (When the drive is non-resonant, there is a quadratic photon term reflecting the splitting due to the finite detuning.)

As first observed in Ref.~\cite{Alsing1991} and later clarified in Ref.~\cite{Alsing1992}, the rotating frame Hamiltonian $\hat{H}_{\textrm{rf}}$ develops an instability as the drive strength $\mathcal{E}$ is increased whereby a discrete spectrum at weak driving gives way to a continuum at strong driving.
This transition occurs for all eigenstates in the spectrum simulataneously and may be understood as arising from the competition between the term $\hat{a}^\dagger \hat{\sigma}_- + \hat{a} \hat{\sigma}_+$, which can be diagonalized by the polaritonic eigenstates from Eq.~\eqref{eqn:JCeigen}, and the term $\hat{a}+\hat{a}^\dagger \propto \hat{X}$, which is proportional to the position operator and has no normalizable eigenbasis (the eigenfunctions of the $\hat{X}$ operator are Dirac delta functions which cannot be properly normalized). 
Accordingly, the discrete, quantized spectrum prevails when $g/\mathcal{E} \ll 1$, while the non-normalizable continuum emerges for $\mathcal{E}/g \gg 1$.

The critical point where the spectrum becomes continuous occurs at the critical drive strength 
\begin{equation}
\mathcal{E}_c = \frac12 g  .
\end{equation}
To see this, note that for $\mathcal{E}< \mathcal{E}_{\rm c}$ the Hamiltonian is diagonalizable and eigenstates can be found exactly~\cite{Alsing1992}. (We rederive this result in Appendix \ref{app:drive-eigen}).
In this regime, the eigenvalues retain their $\sqrt{n}$-like spacing even for $\mathcal{E}\neq 0$ and are given by  \begin{equation}
\label{eqn:DJCeigenvalues}
E^{\textrm{rf}}_{\nu} = \pm \sqrt{n} g \left( 1- \epsilon^2 \right)^{\frac34},
\end{equation}
indexed by the signed quantum number $\nu = 0, \ \pm\sqrt{n}$ introduce above.
Here we introduce the dimensionless drive strength parameter
\begin{equation}
\epsilon = \frac{2\mathcal{E}}{g}.
\end{equation}
As we approach the critical point from below, the effective coupling that controls the level spacing is $g_{\rm eff}=g\left( 1- \epsilon^2 \right)^{3/4}$.
Crucially, for $\epsilon^2 = 1$, the level spacing collapses to zero, and the discrete spectrum condenses into a continuum~\cite{Alsing1992}. 
An alternative description of this transition can be found in Ref.~\cite{Gutierrez-Jauregui2018a}, where the eigenvalue problem is mapped on to that of a charged Dirac particle in both electric and magnetic fields. 
Above the critical drive strength, the Hamiltonian $\hat{H}_{\textrm{rf}}$ is no longer diagonalizable and exhibits dynamical instability.
In this case, the cavity dissipation is crucial, since it ultimately limits the photon number in the absence of detuning.

\subsection{\label{sub:symmetries}Symmetries}
In this section, we discuss the important role of symmetries in our model.
Though the cavity driving provides a preferred reference phase and thus destroys the $U(1)$ symmetry, there is still a remnant anti-unitary $\mathbb{Z}_2$ discrete symmetry associated with the rotating-frame Hamiltonian $\hat{H}_{\textrm{rf}}$ \cite{Alsing1991}.
This corresponding transformation acts jointly on the photon and atom in an anti-unitary fashion and is of the form 
\begin{equation}
    \mathscr{C} = \mathscr{P}\mathscr{T},
\end{equation}
where $\mathscr{P} = e^{\pi i \hat{a}^\dagger \hat{a}}$ is the bosonic parity operator, which implements ``spatial" inversion on the bosonic mode while acting trivially on the atomic degree of freedom, and $\mathscr{T}$ is time-reversal.
Since the two-level system is not a real spin, but a pseudo-spin, $\mathscr{T}$ acts on the atomic degree of freedom through complex conjugation alone.
Hence $\mathscr{T}^2 = +\mathds{1}$, and there is no Kramers degeneracy.
Using the quadrature representation of the bosonic ladder operator as $\hat{a} = \frac{\hat{X} + i \hat{P} }{\sqrt{2}}$, we have
\begin{subequations}
\begin{align}
\mathscr{C}^{-1} i \mathscr{C} &= -i \\
\mathscr{C}^{-1} \hat{X} \mathscr{C} &= -\hat{X},\\
\mathscr{C}^{-1} \hat{P} \mathscr{C} &= \hat{P},\\
\mathscr{C}^{-1} \hat{\sigma}_x \mathscr{C} &= \hat{\sigma}_x, \\
\mathscr{C}^{-1} \hat{\sigma}_y \mathscr{C} &= -\hat{\sigma}_y, \\
\mathscr{C}^{-1} \hat{\sigma}_z \mathscr{C} &= \hat{\sigma}_z.
\end{align}
\end{subequations}
The first of these relations is nothing but the statement that $\mathscr{C}$ is anti-unitary. 

Under these transformations, the rotating-frame Hamiltonian obeys the particle-hole type symmetry 
\begin{equation}
\mathscr{C}^{-1} \hat{H}_{\textrm{rf}} \mathscr{C} = - \hat{H}_{\mathrm{rf}}. 
\end{equation}
This implies that to each eigenstate $|E\rangle$ of $H_{\rm rf}$ with energy $E>0$ there exists an eigenstate $|-E\rangle=\mathscr{C}|E\rangle$ with energy $-E$. For $\mathcal{E}<\mathcal{E}_{\rm c}$, these are, of course, the eigenvalues displayed in Eq.\ \eqref{eqn:DJCeigenvalues}, but the analysis presented here reveals that the particle-hole symmetry also exists for $\mathcal{E}>\mathcal{E}_{\rm c}$. Note that, for $\mathcal{E}=0$, the operator $\mathscr{C}$ relates the two ``polariton branches'' in Eq.\ \eqref{eqn:JCeigen}.
Indeed, using that $\mathscr{C}|n\rangle = (-1)^{n}|n\rangle$ for the noninteracting bosonic Fock states, we see that the transformation inverts the spectrum by switching between the polariton branches for $n>0$ according to $\mathscr{C}|\nu\rangle_0 \propto |-\nu\rangle_0$. The noninteracting vacuum state $|0\rangle$ with $E=0$ is automatically particle-hole symmetric.

We find that, while the transformation leaves the ladder-operators intact (apart from an overall phase),
\begin{align}
    \mathscr{C}^{-1} \hat{a} \mathscr{C} &= -\hat{a},\\
    \mathscr{C}^{-1} \hat{\sigma}_{-} \mathscr{C} &= \hat{\sigma}_-,
\end{align}
it acts non-trivially on coherent states built from them. 
In particular, for a bosonic coherent state 
\begin{align}
|\alpha\rangle =e^{-|\alpha|^2/2}  \sum_n \frac{(\alpha)^{n} }{\sqrt{n!}} |n\rangle,
\end{align}
the conjugated state is $\mathscr{C} |\alpha\rangle = |-\alpha^*\rangle$.

The critical point investigated in this work is associated with a spontaneous breaking of this $\mathbb{Z}_2$ symmetry in the nonequilibrium steady state.
Potential order parameters are the operators $\langle\hat{X}\rangle\propto \text{Re}\langle \hat{a}\rangle$ and $\langle\hat{\sigma}_y\rangle$, both being odd under the transformation $\mathscr{C}$.
In the next section, we compute the steady-state behavior of these quantities using a semiclassical approximation and show that indeed they ought to exhibit nonanalytic behavior at the critical point.

\section{\label{sec:mft}Mean-Field Theory}

In this section, we analyze the master equation \eqref{eqn:lindblad} in terms of a mean-field or semi-classical theory.
In this scheme, we compute the time evolution of the expectation values
\begin{equation}
\begin{aligned}
& \langle \hat{a} \rangle = \mbox{tr} \left( \hat{\rho}(t) \hat{a}\right), \\
& \langle \hat{\vec{\sigma}}\rangle = \mbox{tr} ( \hat{\rho}(t) \hat{\vec{\sigma}})
\end{aligned}
\end{equation}
under the approximation that correlations factorize according to 
\begin{equation}
    \label{eqn:mean-field-approx}
    \langle \hat{a} \hat{\vec{\sigma}} \rangle \approx \langle \hat{a} \rangle \langle \hat{\vec{\sigma}}\rangle.
\end{equation}
Here $\hat{\vec{\sigma}}= (\hat{\sigma}_x, \hat{\sigma}_y,\hat{\sigma}_z)^T $ is the vector of the atomic Pauli matrix operators.
Under this assumption, the equations of motion for the expectation values close, and we have 
\begin{subequations}
\label{eqn:mean-field} 
\begin{align}
& \frac{d\langle \hat{a} \rangle }{dt} = -\kappa\langle \hat{a} \rangle -i (\mathcal{E} + g \langle \hat{\sigma}_{-} \rangle ), \\
& \frac{d\langle \hat{\sigma}_{-} \rangle  }{dt} = ig \langle \hat{a} \rangle \langle \hat{\sigma}_z \rangle,\\
& \frac{d\langle \hat{\sigma}_z \rangle }{dt} = -2ig \Bigl(\langle \hat{a} \rangle \langle \hat{\sigma}_{-} \rangle ^* - \langle \hat{a} \rangle^* \langle \hat{\sigma}_{-} \rangle\Bigr). 
\end{align}
\end{subequations}
Notably, these equations conserve the length of the atomic Bloch vector 
\begin{align}
\ell^2 = |\langle \hat{\vec{\sigma}} \rangle| ^2 \leq 1.
\end{align}
Since this quantity is conserved, we have a one-parameter family of steady states labeled by the length of the Bloch vector $\ell$. 
This multiplicity is not expected to survive once we include fluctuations, and is most likely an artefact of the approximation in Eq.~\eqref{eqn:mean-field-approx}.

The steady-state solution of equations~\eqref{eqn:mean-field} is obtained by setting the time-derivatives to zero and solving the resulting algebraic equations. 
For a fixed value of $\ell^2$, the solution to these equations depends on whether the drive is above or below the critical value of $\epsilon_{\rm c}^2 = \ell^2$. In the following, we examine these solutions in more detail. A summary of our findings is presented in Tab.~\ref{tab:steady-state}.

For $\epsilon^2 < \ell^2$ (below the critical drive strength), the solution to the steady-state equations is  
\begin{align}
& \langle \hat{a} \rangle = 0, \\ 
 \label{eqn:sigma-minus} & \langle \hat{\sigma}_{-} \rangle = -\frac12 \epsilon, \\
& \langle \hat{\sigma}_z \rangle= \pm \sqrt{\ell^2- \epsilon^2 }. 
\end{align}
In this case, we may interpret the atomic Bloch vector $\langle \hat{\vec{\sigma}}\rangle$ as a dipole re-radiating a coherent field which self-consistently cancels the externally imposed driving field, producing a total photon field of $\langle \hat{a}\rangle =0$.
This steady state corresponds to the ``disordered" phase with respect to the $\mathbb{Z}_2$ symmetry induced by $\mathscr{C}$. Indeed, the expectation values
\begin{align}
& \langle \mathscr{C}^{-1} \hat{X}\mathscr{C} \rangle = -  \langle \hat{X} \rangle, \\
& \langle \mathscr{C}^{-1} \hat{\sigma}_y \mathscr{C} \rangle = -  \langle\hat{\sigma}_y \rangle,
\end{align}
which transform non-trivially under $\mathscr{C}$, are both zero in this regime.
Importantly, the solution is independent of $\kappa$ (or, more precisely, independent of the dimensionless parameter $g/2\kappa$). 
Furthermore, a simple analysis of the non-linear equations of motion around this steady state reveals that the solution with $\langle \hat{\sigma}_3 \rangle <0$ is a dynamically stable fixed point, while $\langle \hat{\sigma}_3 \rangle > 0$ is dynamically unstable (see Appendix~\ref{app:stability} for the analysis).

For $\epsilon^2 > \ell^2$ (above the critical point), we find the steady-state solution to form a new pair of dynamical fixed points, given by
\begin{align}
   \langle \hat{a} \rangle &= \frac{g}{2\kappa}\sqrt{\epsilon^2 -\ell^2} \Bigl(\mp \frac{\ell}{\epsilon} -i\sqrt{1-\frac{\ell^2}{\epsilon^2}} \Bigr),\\ 
 \label{ordermf}      \langle \hat{\sigma}_{-} \rangle &= -\frac{\ell^2}{2\epsilon} \mp i \frac{\ell}{2} \sqrt{1-\frac{\ell^2}{\epsilon^2}}, \\
     \langle \hat{\sigma}_z \rangle &= 0.
\end{align}
The two solutions labelled by ``$\pm$"  are related by the action of $\mathscr{C}$. 
In these steady states, both $\textrm{Re} \langle \hat{a}\rangle$ and $\langle \hat{\sigma}_y\rangle $ are nonzero, thereby spontaneously breaking the $\mathbb{Z}_2$ symmetry. 
The order parameter
\begin{align}
    \textrm{Re}\langle \hat{a} \rangle = -\frac{g}{2\kappa} \langle \hat{\sigma}_y \rangle 
\end{align}
characterizes the continuous nonequilibrium phase transition. 
Let us remark that, in contrast to the disordered phase, the mean-field steady state for $\epsilon^2 > \ell^2$ does depend on the value of $\kappa$, with the mean-field bosonic order parameter $\langle \hat{X}\rangle$ diverging as $\kappa \rightarrow 0$.
Interestingly, there is precedent for the occurrence of $\mathcal{PT}$-symmetry breaking transitions in non-Hermitian dynamics of classical spins~\cite{Galda17}.

\renewcommand{\arraystretch}{1.5}
\begin{table*}[t!]
\centering
\begin{tabular}{| c | c | c | c | c| }
\hline
Phase & Symmetry & Photon Field & Atomic Bloch Vector & \ Linear Stability \  \\
\hline
\hline
\ $\epsilon^2 < \ell^2$ \   & \ $\mathbb{Z}_2$-disordered \  & $\langle \hat{a} \rangle = 0$ &
\begin{tabular}{c}
    $\langle \hat{\sigma}_{-} \rangle = -\frac12 \epsilon$ \\
    $ \langle \hat{\sigma}_3 \rangle = \pm \sqrt{\ell^2- \epsilon^2 }$ \\
\end{tabular} & 
\begin{tabular}{c}
 ``$-$'' stable, \\ ``$+$'' unstable 
\end{tabular}\\
\hline
$\epsilon^2 > \ell^2$ & $\mathbb{Z}_2$-ordered & \ $\langle \hat{a} \rangle = \frac{g}{2\kappa}\sqrt{\epsilon^2 -\ell^2} \left(\mp \frac{\ell}{\epsilon} -i\sqrt{1-\ell^2/\epsilon^2} \right)$ \ & 
\begin{tabular}{c} $\ \langle \hat{\sigma}_{-} \rangle = -\frac{\ell^2}{2\epsilon} \mp i \frac{\ell}{2} \sqrt{1-\ell^2/\epsilon^2}$ \ \\ $\langle \hat{\sigma}_3 \rangle = 0$ \end{tabular}
& neutrally stable \\
\hline
\end{tabular}
\caption{Summary of semi-classical steady-state solutions for the photon field and atomic Bloch vector from Eqs.\ \eqref{eqn:mean-field}. For a fixed value of $\ell$, we find two phases separated by a critical point at $\epsilon^2 = \ell^2$. The associated order parameter is $\text{Re}\langle \hat{a}\rangle = -\frac{g}{2\kappa}\langle\hat{\sigma}_y\rangle$.}
\label{tab:steady-state}
\end{table*}
\renewcommand{\arraystretch}{1}

\begin{figure}
    \centering
    \includegraphics[width=\linewidth]{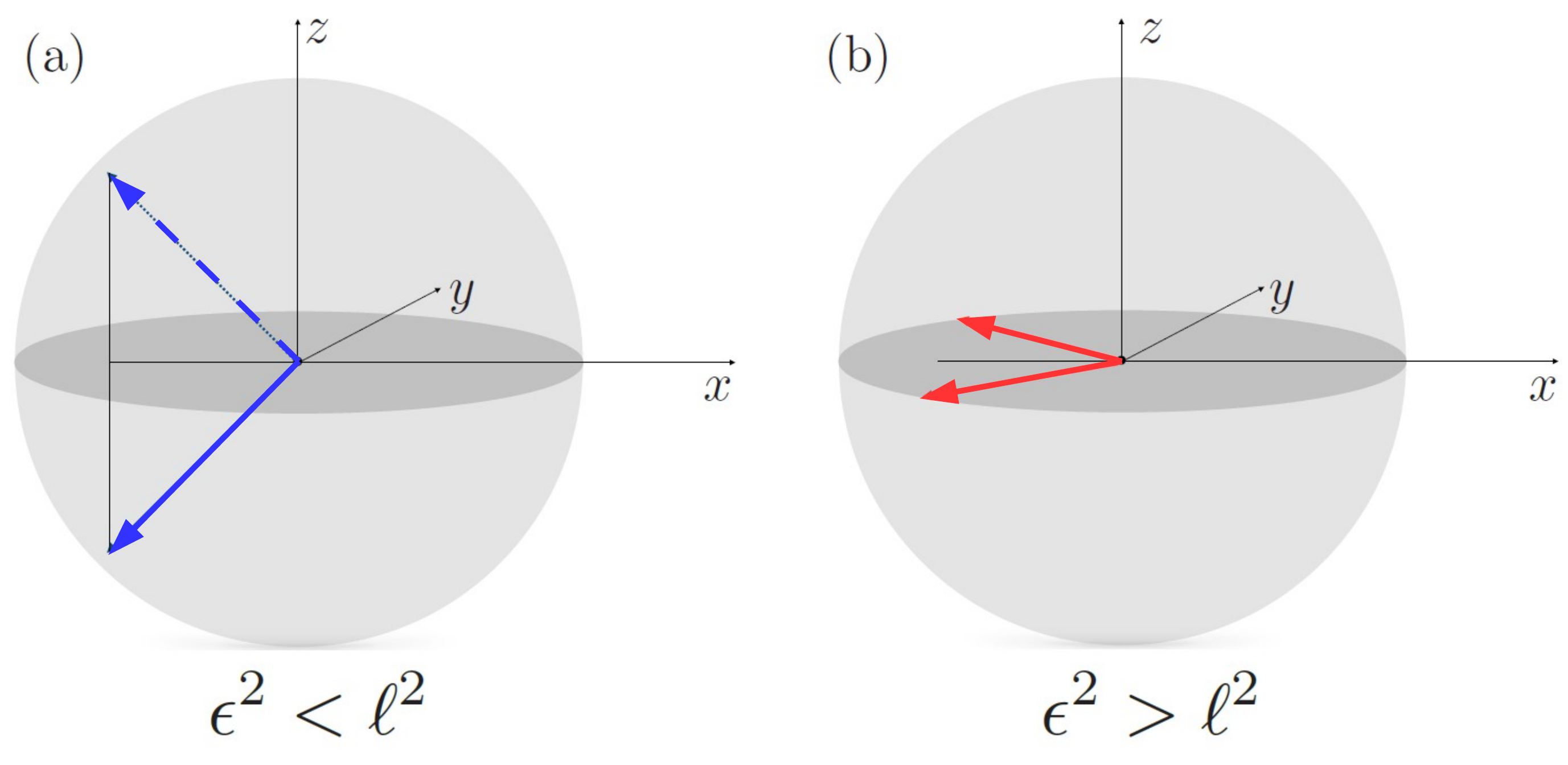}
    \caption{Representation of the semi-classical steady-state solution in terms of the atomic Bloch vector $\langle \hat{\vec{\sigma}}\rangle$. (a) $\mathbb{Z}_2$-disordered phase for $\epsilon^2 < \ell^2$. 
    The solid blue vector corresponds to $\langle \hat{\sigma}_z\rangle = -\sqrt{\ell^2-\epsilon^2}$ and is dynamically stable; the dashed one,  $\langle \hat{\sigma}_z\rangle =+\sqrt{\ell^2-\epsilon^2}$, is unstable.
    (b) $\mathbb{Z}_2$-ordered phase for $\epsilon^2 > \ell^2$.
    Both fixed points with $\langle \hat{\sigma}_z\rangle =0$ in Eq. \eqref{ordermf} are neutrally stable.}
    \label{fig:bloch-vectors}
\end{figure}

The non-analyticity of the steady state in this zero-dimensional system is not related to the more common thermodynamic limit in which the system size diverges. 
Rather, as was clarified in Ref. \cite{Carmichael2015}, the thermodynamic limit here corresponds to $\kappa\to 0^+$, or, equivalently, diverging excitation numbers~\cite{Rota2019,Paz,Casteels2017}. 
While a finite value of $\kappa>0$ limits the population of a single mode, a system without dissipation may exhibit a phase transition through a diverging population of certain modes. 
As such, for finite $\kappa$, we expect the system to exhibit finite-size crossover behavior instead of a true non-analytic phase transition.
We simplify matters by exclusively focusing on the $\kappa\rightarrow 0^+$ limit; however, studying the effects of finite-size fluctuations~\cite{Paz} due to a small, but finite $\kappa$ would be an interesting subject for future investigations.  
Bearing this in mind, we proceed on to study the $\kappa\to 0$ limit in the following section by mapping the quantum master equation on to a classical rate equation.

\section{\label{sec:rateeqn}Rate Equation}

We now proceed to derive a rate equation governing the long-time behavior of the Lindblad equation Eq.~\eqref{eqn:lindblad} in the rotating frame. This analysis is restricted to the regime below the critical point, so we only consider $\epsilon< \epsilon_{\rm c}$ in this section.

\subsection{Mapping to Classical Rate Equation}

\begin{figure}
    \centering
    \includegraphics[width=\linewidth]{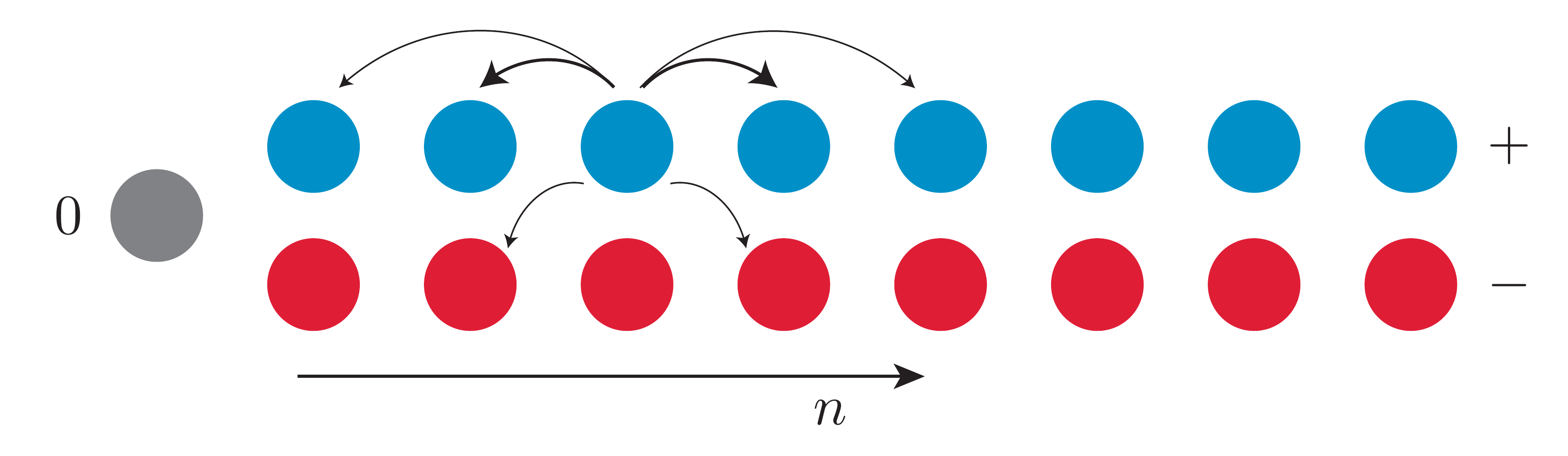}
    \caption{Fock-space lattice of eigenstates $|\nu\rangle$ of $\hat{H}_{\rm rf}$ labeled by quantum number $\nu = s\sqrt{n}$.
    In this arrangement the system is organized as a semi-infinite one-dimensional lattice with two parallel excitation pathways.
    The lattice sites are indexed by $n=0,1,2,...$, which plays the role of a ``spatial" variable, and by  $s= +1$ (blue) or $s=-1$ (red), indicating the branch.
   States that are symmetric under the transformation $\mathscr{C}$ are characterized by an equal distribution function on both branches.
   We map the rate equation \eqref{eqn:dissipator} to the problem of hopping on this lattice, schematically depicted here for a generic site by the arrows emanating from the site.}
    \label{fig:state-lattice}
\end{figure}

Recall from the exact diagonalization of $\hat{H}_{\rm rf}$ that, below the critical driving strength, the Hamiltonian which governs the coherent part of the evolution, has a discrete spectrum. We label the associated dressed energy levels by the single quantum numbers $\mu = r\sqrt{m},  \nu=s\sqrt{n}$ with $r,s = \pm$ and $m,n = 0,1,2...$.
We work in units where $g = 1$.
The eigenenergy from Eq.\ \eqref {eqn:DJCeigenvalues} is then given by 
\begin{equation}
    E_{\mu} = (1-\epsilon^2)^{\frac34} \mu .
\end{equation}
The spacing between two levels is finite and reads
\begin{equation}
| E_{\mu} - E_{\nu} | =(1-\epsilon^2)^{\frac34} |\mu - \nu| \geq 0 ,
\end{equation}
with equality if and only if $\nu = \mu$. 
The eigenstates $|\mu\rangle, |\nu\rangle$ are known in closed form and given in Appendix~\ref{app:drive-eigen}.
It is helpful to visualize the quantum states as forming a lattice in Fock space, see Fig.~\ref{fig:state-lattice}.

Working in the energy basis simplifies the quantum master equation to the form 
\begin{equation}
\label{eqn:DDJCMenergy}
\frac{\partial \rho_{\mu\nu}}{\partial t} = - i (E_\mu - E_\nu)\rho_{\mu\nu} + \kappa \sum_{\alpha\beta}\mathcal{K}_{\mu\nu}^{\alpha\beta} \rho_{\alpha\beta}, 
\end{equation}
where $\rho_{\mu\nu} = \langle \mu | \hat{\rho} | \nu \rangle $ and 
\begin{equation}
\label{eqn:dissipator}
\mathcal{K}_{\mu\nu}^{\alpha\beta} = 2\langle \mu| \hat{a}|\alpha\rangle \langle \beta | \hat{a}^\dagger |\nu\rangle - \langle \mu| \hat{a}^\dagger \hat{a} |\alpha\rangle \delta_{\nu\beta} - \langle \beta| \hat{a}^\dagger \hat{a} |\nu\rangle \delta_{\mu\alpha}.
\end{equation}
In the following, we fix $0<\epsilon<1$ to some value close to unity, below the transition point at $\epsilon_{\rm c}^2 = 1$. 
Then we find the steady state by taking $\kappa\rightarrow 0^+ $ for fixed $\epsilon$ by considering the limit 
\begin{align}
\label{eqn:rhoss} \rho_{\textrm{ss}}(\epsilon)  = \lim_{\kappa \rightarrow 0^+}\lim_{t\rightarrow \infty} \rho(t).
\end{align}
Here $\rho(t)$ obeys Eq.~\eqref{eqn:lindblad}.

To lowest order in $\kappa$, the steady-state solution is given by a diagonal density matrix. 
Indeed, Eq.\ \eqref{eqn:DDJCMenergy} reads $(E_\mu-E_\nu)\rho_{\mu\nu}=0$ in this case. 
Hence, we find the first nontrivial correction due to $\kappa$ to be given by the \emph{classical} master equation
\begin{equation}
\label{eqn:rate}
\frac{\partial \rho_\nu }{\partial t} =  2\kappa \sum_{\mu} \Gamma_{\nu\mu}\rho_\mu
\end{equation}
for the diagonal entries $\rho_\mu\equiv \rho_{\mu\mu}$ of the density matrix. The transition rates read
\begin{equation}
\Gamma_{\nu\mu} \equiv \frac12 \mathcal{K}_{\nu\nu}^{\mu\mu} = | \langle \nu |\hat{a} | \mu\rangle|^2 - \delta_{\mu\nu} \langle \mu | \hat{a}^\dagger \hat{a} | \mu\rangle.
\end{equation}
Note that the diagonal elements are fixed by the sum rule, which ensures the master equation to be trace preserving, $\partial_t (\sum_\nu \rho_\nu)=0$, so that
\begin{align}
 \label{eqn:trace} \sum_\nu\Gamma_{\nu\mu}=0:\ \Gamma_{\mu\mu} = - \sum_{\nu \neq \mu}  \Gamma_{\nu\mu}.
\end{align}
Starting from Eq.\ \eqref{eqn:rate}, it is then evident that we need to solve the linear algebraic problem 
\begin{align}
\label{eqn:ss-eigen-prob}
\sum_\mu \Gamma_{\nu\mu} \rho_{\textrm{ss},\mu}(\epsilon) =0.
\end{align}
Up to an overall equilibration time-scale defined by $\kappa$, the transition-rate matrix $\Gamma$ only depends on $\epsilon $. Consequently, the critical behavior of the steady state in the thermodynamic limit is a function of the control parameter $\epsilon$.

Using the rate equation, it is easy to see why $\kappa\to 0^+$ controls the number of excitations.
The rate equation description only applies to those levels which are well separated as compared to the decay rate $\kappa$.
The spacing between adjacent levels decreases with increasing $n$, and for a given maximal $n\leq N_{\textrm{max}}$, the smallest spacing is 
\begin{align}
 \nonumber |E_{\nu} - E_{\mu}| & \geq (1-\epsilon^2)^{\frac34}|\sqrt{N_{\textrm{max}}} - \sqrt{N_{\textrm{max}}-1}|\\
 &\sim \frac{(1-\epsilon^2)^{\frac34}}{2\sqrt{N_{\textrm{max}}}}.
\end{align}
This must be much larger than $\kappa$ in order to justify throwing away the off-diagonals in the density matrix. 
This provides us with the unitless figure of merit 
\begin{align}
\frac{(1-\epsilon^2)^{\frac34}}{2\sqrt{N_{\textrm{max}}}} \gg \kappa \Rightarrow N_{\textrm{max}} \ll \left(\frac{(1-\epsilon^2)^{\frac34}}{2\kappa}\right)^2 .
\end{align}
We interpret this $N_{\rm max}$ as the effective system size, since our description will encounter finite-size quantum fluctuations (due to the reappearance of the quantum coherence) when the typical quantum numbers $n$ are of order $N_{\rm max}$. 
Thus, the thermodynamic limit is taken by first fixing $\epsilon^2 <1$ so the numerator is finite, and then taking $\kappa \to 0$ so that $N_{\rm max}$ is effectively unconstrained. 
This is precisely the limit described above and constitutes the parameter range considered in the following sections.

\subsection{Analysis of Transition Rates}

The aim of this section is to understand those properties of the transition rates $\Gamma_{\nu\mu}$ that govern the criticality of the steady-state solution. In particular, we perform an asymptotic expansion of the rates for large $\nu+\mu$, which gives a very accurate approximation to the full expressions $\Gamma_{\nu\mu}$, and, at the same time, allows us to solve the rate equation.

Using the exact expression for the eigenstates presented in  Appendix~\ref{app:drive-eigen}, we find that the matrix elements take the form 
\begin{equation}
\label{eqn:transition-rates}
    \Gamma_{\nu\mu} = |\langle n,s | (u\hat{a} + v\hat{a}^\dagger) | m,r\rangle |^2,
\end{equation}
with $\nu=s\sqrt{n}$, $\mu=r\sqrt{m}$, $u = \cosh\eta,\ v=\sinh\eta$, and
\begin{align}
    \eta = -\frac{1}{4} \ln(1-\epsilon^2).
\end{align} 
As $\epsilon\to 1$, we have $\eta\to \infty$, leading to a divergence of the  Bogoliubov coefficients $u, v \sim e^\eta/2$ at the critical point. In contrast, the matrix elements $\langle n,s|\hat{a}|m,r\rangle$ remain finite at the transition. Therefore, we will henceforth set $\epsilon=1$ in computing the matrix elements and only take into account the leading divergences of $u$ and $v$.

\begin{figure}
    \includegraphics[width=\linewidth]{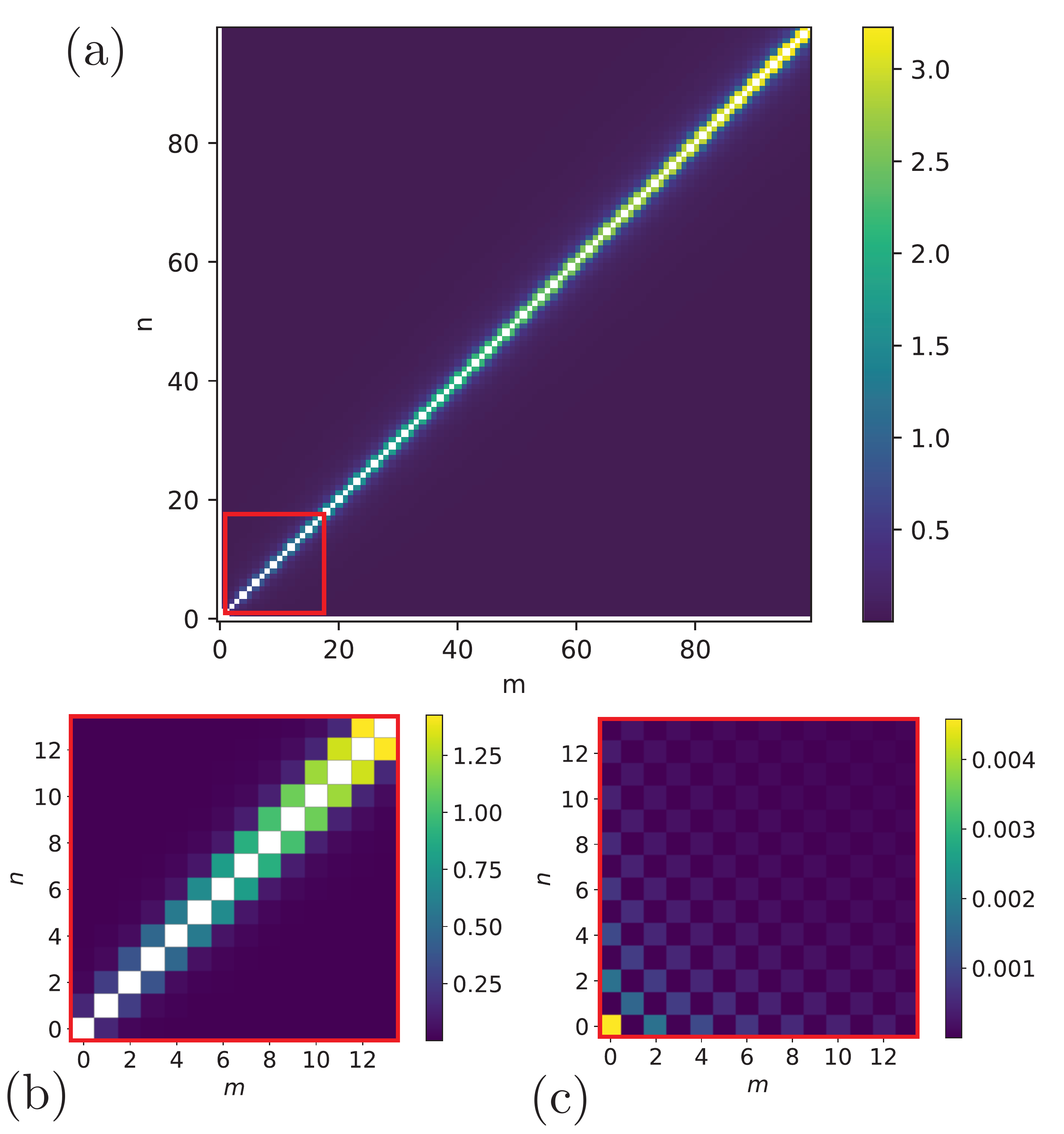}
    \caption{These plots shows that interbranch contributions to the symmetric matrix element $\Gamma_{\nu\mu}+\Gamma_{\mu\nu}$  are negligible. (a) Intrabranch matrix element $S_{nm}$. (b) Close-up on $|S_{nm}|^2$ in the subregion framed red.
    (c) Interbranch $|S^{(\rm inter)}_{nm}|^2$, defined in the main text, shown for the same region. We observe interbranch transitions to be several orders of magnitude smaller than intrabranch transitions and to display a fast decay for $(m, n) \neq (0,0)$.}
    \label{fig:symmetric-compare}
\end{figure}

The transition rates respect the $\mathbb{Z}_2$ symmetry corresponding to $\mathscr{C}$ (which relates the two excitation branches $\pm \sqrt{n}$ to each other).
Thus,
\begin{align}
    \Gamma_{\nu\mu} = \Gamma_{-\nu,-\mu}.
\end{align}
Furthermore, at large excitation number $n$, the rate for ``interbranch" transitions [$\text{sign} (\nu) = -\text{sign} (\mu)$)] decays as $\sim 1/n$, see Fig.~\ref{fig:symmetric-compare}.
We can therefore restrict our attention to only those processes which induce ``intrabranch" transitions [$\text{sign} (\nu) = \text{sign} (\mu)$].
Within this well-justified approximation, the steady-state density matrix will also respect the $\mathbb{Z}_2$ symmetry and can be written as
\begin{equation}
    \label{eqn:steady-state-branch}
    \rho_{\textrm{ss}} = \sum_{\nu} \rho(n) |\nu\rangle \langle \nu|.
\end{equation}
Recall that $n = \nu^2\Leftrightarrow \nu = \pm \sqrt{n}$, so that the steady state contains an equal mixture of both positive and negative energy states.

Characterizing the steady state amounts to determining the population function $\rho(n)$. 
We set $r=s=+1$ and write
\begin{align}
    \Gamma_{nm} = \Gamma_{\nu\mu}|_{s=r=1}
\end{align}
in a slight abuse of notation. 
The behavior of the steady-state solution $\rho(n)$ is completely determined by the control parameter $\epsilon$. 
From Eq.\ \eqref{eqn:ss-eigen-prob}, we find
\begin{align}
  \label{eqn:rho-ss-2}  \sum_m \Gamma_{nm} \rho(m)=0,
\end{align}
and so the steady-state population $\rho(m)$ is a \emph{right-nullvector} of the transition rate matrix $\Gamma=(\Gamma_{nm})$.

We now discuss the behavior of the rates close to criticality. 
The leading  divergence for $\eta\to \infty$ follows from
\begin{align}
\nonumber \Gamma_{nm}  ={}&  \Bigl| e^{\eta}\langle n,+ | \frac{\hat{a}+\hat{a}^\dagger}{2}| m,+\rangle + e^{-\eta}\langle n,+ | \frac{\hat{a}-\hat{a}^\dagger}{2}| m,+\rangle \Bigr|^2\\
 \nonumber \sim{}& e^{2\eta}\Bigl| \Bigl\langle n,+\Bigl| \frac{\hat{a}+\hat{a}^\dagger}{2}\Bigr| m,+\Bigr\rangle\Bigr|^2  \\
 &\times \Biggl( 1 + 2e^{-2\eta}\frac{ \langle n,+ | \left(\hat{a}-\hat{a}^\dagger\right)| m,+\rangle}{\langle n,+ | \left(\hat{a}+\hat{a}^\dagger\right)| m,+\rangle}\Biggr). \end{align}
 In the second equality, we have utilized the fact that the matrix elements of $\hat{a}$ in the eigenbasis $|n,+\rangle$ are purely real.
We introduce the symmetric and anti-symmetric matrix elements 
\begin{equation}
    \begin{aligned}
    & S_{nm} = S_{mn} = \Bigl\langle n,+\Bigl| \frac{\hat{a}+\hat{a}^\dagger}{2}\Bigr| m,+\Bigr\rangle, \\
    & B_{nm} = - B_{mn} = \frac{\langle n,+ | \hat{a}-\hat{a}^\dagger| m,+\rangle}{\langle n,+|\hat{a}+\hat{a}^\dagger| m,+\rangle}. \\
    \end{aligned}
\end{equation}
Up to an overall prefactor, which can be absorbed into a rescaling of time, we can then write the transition rate matrix as 
\begin{equation}
    \label{eqn:rate-matrix}
    \Gamma_{nm} = |S_{nm}|^2 \left( 1 + p B_{nm}\right)
\end{equation}
with asymmetry parameter 
\begin{equation}
    p = 2e^{-2\eta} = 2\sqrt{1-\epsilon^2}.
\end{equation}

A nonzero value of $p$ implies $\Gamma_{nm}\neq \Gamma_{mn}$. If $p=0$, then $\Gamma$ is symmetric and the constant state $(1,1,\dots,1)$ is a nullvector. (This follows from the trace preserving property discussed in Eq.\ \eqref{eqn:trace}, which implies $\sum_n\Gamma_{nm}=0$.) Such a state, however, is non-normalizable in an infinite Hilbert space, where $n$ is allowed to be arbitrarily large. Thus, the small asymmetry induced by $p >0$ is crucial in determining the width of the steady state in the thermodynamics limit, and we expect the width to diverge as $p\rightarrow 0$.

The matrix elements $S_{nm}$ and $B_{nm}$ can be computed in closed form, see Appendix~\ref{app:mat-el}.
However, they are too complicated to be useful in practice.
Therefore, in order to simplify the rate equation, we consider their asymptotic behavior for $n+m\to \infty$ while keeping $n-m$ fixed. We then find 
\begin{equation}
\label{eqn:s-b-scaling}
\begin{aligned}
&   S_{nm} \sim \sqrt{n+m}\ | n-m|^{-\frac53},  \\
&   B_{nm} \sim  |n-m|^{\frac13}\ \textrm{sign}(m-n). \\    
\end{aligned}
\end{equation}
This scaling turns out to be a good approximation even for moderate values of $n+m$. In the following, we sketch how these forms are obtained analytically, and corroborate them numerically using the full expressions.

The analytical derivation of Eq.\ \eqref{eqn:s-b-scaling} is presented in detail in Appendix \ref{app:mat-el}. The central steps are the following. Define the matrix element 
\begin{equation}
    A_{nm} = \langle n,+| \hat{a} | m,+\rangle,
\end{equation}
which can be expressed in terms of associated Laguerre polynomials. Now set $m=n+\delta$ and consider the limit $n\to \infty$ while keeping $\delta\in\mathbb{Z}$ fixed. Using the asymptotic behavior of Laguerre polynomials, we find, for $\delta\neq 0$, that
\begin{align}
    A_{n,n+\delta} \sim \sqrt{n}\  \frac{1}{\delta}J_{\delta-1}(\delta),
\end{align}
with $J_k(y)$ the Bessel function of the first kind. This implies that the symmetric and anti-symmetric matrix elements behave as
\begin{align}
\label{eqn:s-b-bessel 2}    S_{n,n+\delta} &\sim \sqrt{n} \frac{1}{2|\delta|}\Bigl( J_{|\delta|-1}(|\delta|) - J_{|\delta|+1}(|\delta|) \Bigr) ,  \\
\label{eqn:s-b-bessel}    B_{n,n+\delta} &\sim \text{sign}(\delta) \frac{ J_{|\delta|-1}(|\delta|) + J_{|\delta|+1}(|\delta|)}{J_{|\delta|-1}(|\delta|) - J_{|\delta|+1}(|\delta|)}.    
\end{align}
For $\delta\to \infty$, the Bessel functions can be expanded in terms of Airy functions, giving a simple power-law behavior in $\delta$ as $S_{n,n+\delta} \sim C_1\sqrt{n} |\delta|^{-5/3}$ and $B_{n,n+\delta}\sim C_2 |\delta|^{1/3}$ with two constants $C_{1,2}$. This is actually a very good approximation even for moderate values of $\delta$ of order unity. We have $C_2\approx 1$. After another rescaling of time, we eventually arrive at  Eq.\ \eqref{eqn:s-b-scaling}.

We now numerically test the simplifications that lead to Eq.\ \eqref{eqn:s-b-scaling} against the exact matrix elements $\Gamma_{\nu\mu}$. In Fig.~\ref{fig:symmetric-compare}, we show the exact result for the symmetric matrix elements for intrabranch and interbranch transitons, $S_{nm} \propto (\Gamma_{\nu\mu}+\Gamma_{\mu\nu})_{r=s=1}$ and $S_{nm}^{(\rm inter)} \propto (\Gamma_{\nu\mu}+\Gamma_{\mu\nu})_{r=-s=1}$, respectively. This justifies our approximation to neglect the interbranch transition rates, which, as can be seen clearly in the plot, are several orders of magnitude smaller than the intrabranch rates.

\begin{figure}
    \centering
    \includegraphics[width=\linewidth]{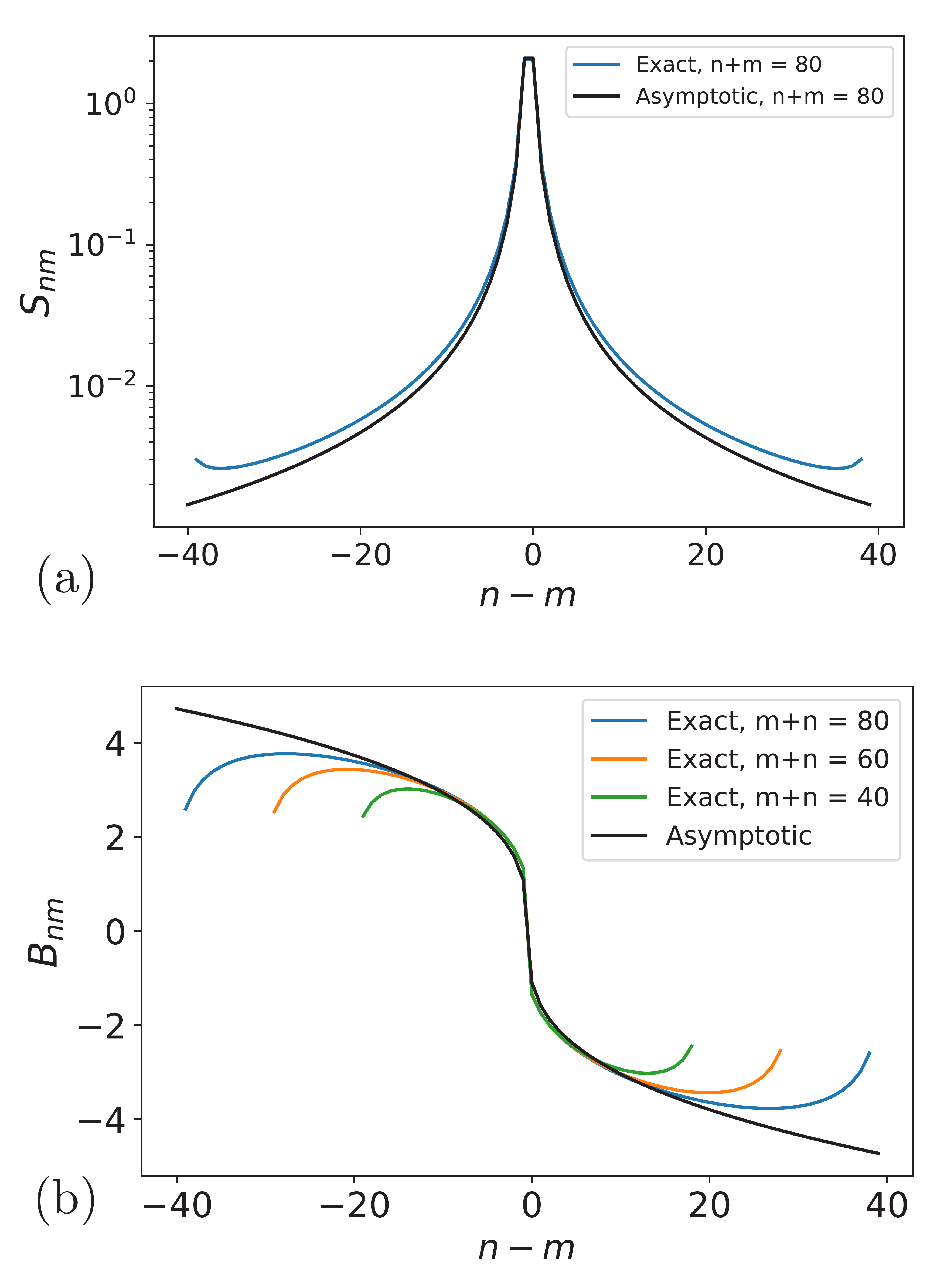}
    \caption{(a) Exact $S_{nm}$ and its asymptotic behavior from Eq.~\eqref{eqn:s-b-scaling} for $n+m = 80$. 
    Note the logarithmic scale.
   (b) Exact $B_{nm}$ and corresponding asymptotic function from Eq.~\eqref{eqn:s-b-scaling} for various choices of $n+m$.
    Note that the asymptotic expression for $B_{nm}$ does not depend on $n+m$, and, indeed, near the main diagonal, all the curves collapse. 
    Deviations between the exact and asymptotic forms are only visible when $|n-m|$ becomes of order $\frac{n+m}{2}$, where the expansion breaks down. }
    \label{fig:slices}
\end{figure}

In Fig.~\ref{fig:slices}, we plot the exact and asymptotic forms of the intrabranch rates $S_{nm}$ and $B_{nm}$ for fixed values of $n+m$. We verify a clear power-law decay of $S_{nm}$, which is tightly localized around the main diagonal.
Obviously, the asymptotic formula becomes worse as $|n-m|$ approaches $\frac{n+m}{2}$, but is still surprisingly accurate. Furthermore, in this region, the magnitude $|S_{nm}|\sim 10^{-3}$. This needs to be compared to values of order unity on the main diagonal. The outer regions are thus unimportant in comparison to transitions which occur near the main diagonal.

Having established the behavior of the symmetric part of the matrix, we now turn to the anti-symmetric part $B_{nm}$.
By construction, we have $B_{nm} \sim \textrm{sign}(m-n)$. In  Fig.~\ref{fig:slices}, we plot slices of $B_{nm}$ for constant $n+m$ as a function of $|n-m|$.
Apart from the anti-symmetry, we see that all slices have the same universal behavior near the main diagonal, given by the power law from Eq.~\eqref{eqn:s-b-scaling}. 
Again, deviations increase as we move further away from the main diagonal, but in this region the function $|S_{nm}|^2$ is small and thus these deviations are not important for the solution of the master equation.

\subsection{Solving the Rate Equation}

\begin{figure}
    \centering
    \includegraphics[width=\linewidth]{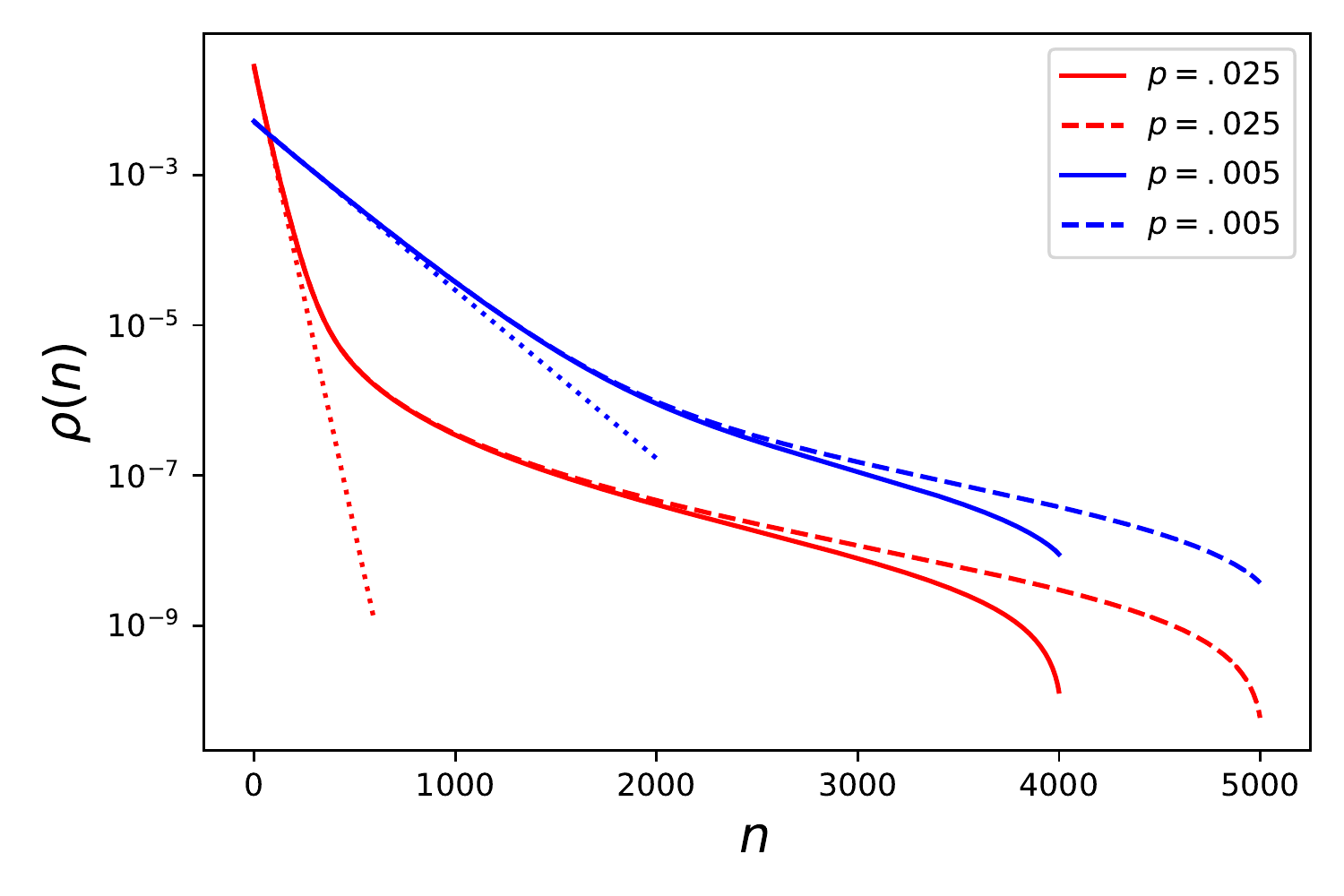}
    \caption{Nonequilibrium steady state below criticality, parametrized through the distribution $\rho(n)$  in Eq.\ \eqref{eqn:steady-state-branch}. 
    We numerically obtain $\rho(n)$ by solving Eq.\ \eqref{eqn:rho-ss-2}, using the  approximations \eqref{eqn:rate-matrix-form} for the transition rates.
    The solid line corresponds to a truncated Fock space with $n\leq N_{\rm max}=4000$, while the dashed line is for $N_{\rm max}= 5000$.
    Both solutions agree well for small values of $n$.
    The straight dotted lines indicate that the initial portion of the distribution is well-approximated by an exponential distribution, independent of $N_{\rm max}$.
    The universal part of $\rho(n)$ only depends on the asymmetry parameter $p=2\sqrt{1-\epsilon^2}$, which controls the distance from the critical point. As $p\to 0$, the distribution becomes broader, occupying more of the highly excited Fock states.}
    \label{fig:steady-state-solutions}
\end{figure}

In the previous section, we have established, both analytically and numerically, that the transition-rate matrix can be very accurately approximated by (for $n\neq m$)
\begin{equation}
\label{eqn:rate-matrix-form}
    \Gamma_{nm} = \frac{n+m}{2}|n-m|^{-\frac{10}{3}} \Bigl( 1 + p |n-m|^{\frac13} \textrm{sign}(m-n) \Bigr),
\end{equation}
with the asymmetry parameter $p = 2e^{-2\eta} = 2\sqrt{1-\epsilon^2}$ controlling the approach to the critical point. [The diagonal elements $\Gamma_{nn}$ are fixed by the sum rule Eq.~\eqref{eqn:trace}.] 
We numerically solve Eq.\ \eqref{eqn:rho-ss-2} using this form of $\Gamma_{nm}$ for small values of $p$.
Specifically, we find the steady-state population $\rho(n)$ as the right-eigenvector of the matrix $\Gamma$ with eigenvalue zero.
By virtue of the approximations made in the transition-rate matrix, we are able to compute the matrix elements for large values of $n\leq N_{\rm max}$. 
The result of this analysis is shown in Fig.~\ref{fig:steady-state-solutions}.

\begin{figure}
    \centering
    \includegraphics[width=\linewidth]{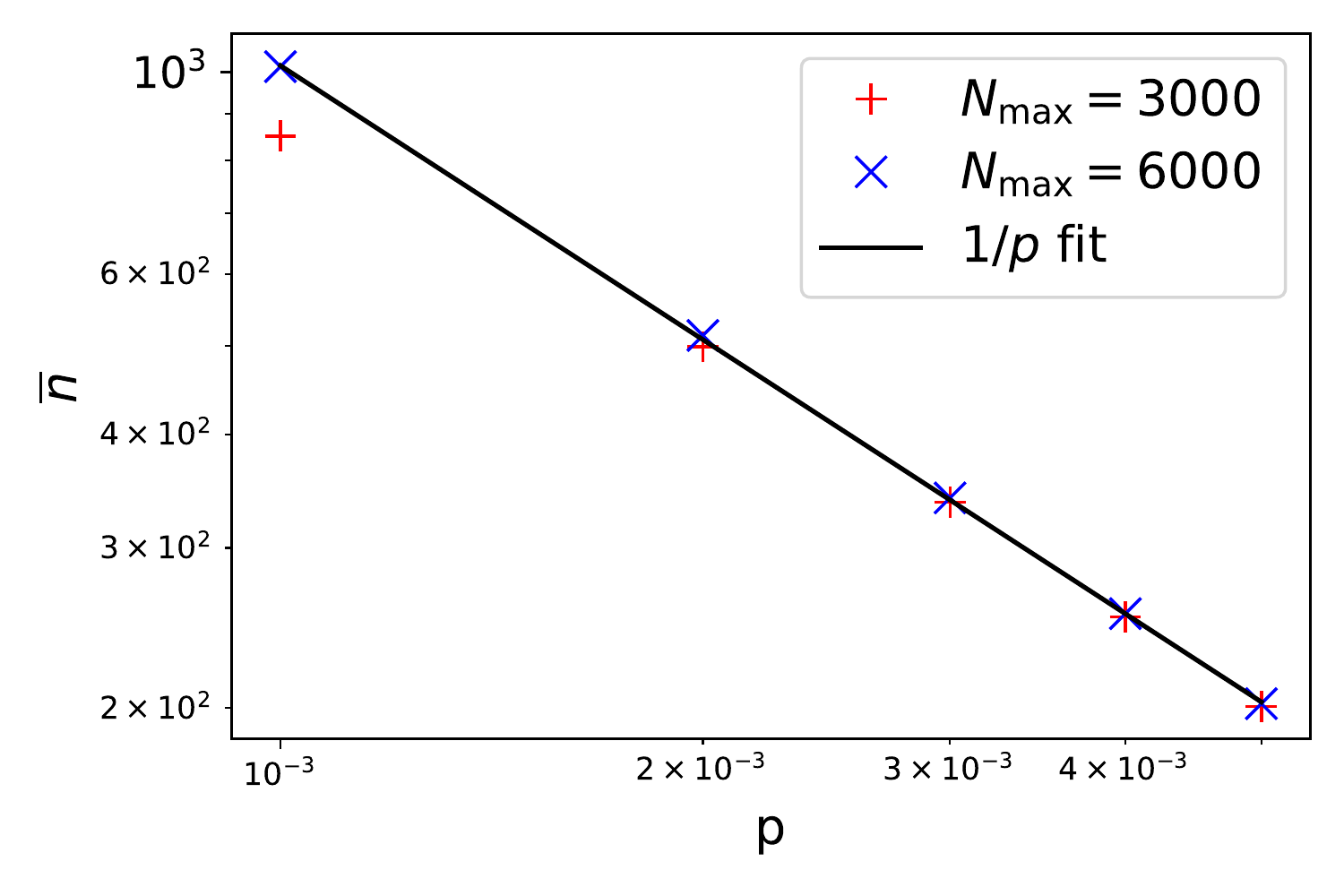}
    \caption{Average quantum number $\bar{n}$ in the steady state vs. asymmetry parameter $p$, obtained by numerically solving the rate equation with transition rate matrix~\eqref{eqn:rate-matrix-form}. 
    The data is shown for two different Fock space cutoffs (such that $n\leq N_{\textrm{max}}$).
    The scaling behavior is very well-described by $\bar{n} \sim 1/p$ (black line).
    For larger $N_{\rm max}$, the agreement becomes better, with deviations visible only at the smallest $p$.}
    \label{fig:nbar-scaling}
\end{figure}

As we decrease $p\rightarrow 0$, the distribution $\rho(n)$ becomes wider.
The decay in $n$ is roughly exponential, which we will later support by a simple qualitative argument. 
The steady-state distribution can be characterized by its spread in terms of the mean quantum number 
\begin{equation}
    \bar{n} = \sum_n n \rho(n) .
\end{equation}
Note that, since we work in the dressed basis, this is not the average photon number, but a closely related quantity, as we detail below. 
If $\rho(n)$ decays exponentially over a length scale $\xi$, then $\bar{n}\sim \xi$. (This may also be taken as a definition of $\xi$.)
As $p \rightarrow 0$, the distribution becomes wider and $\bar{n}$ diverges.
In Fig.~\ref{fig:nbar-scaling}, we plot $\bar{n}$ for small values of $p$ for two different cutoffs $N_{\rm max}$.
For sufficiently large system size, we observe $\bar{n}\sim 1/p$.

To summarize, we have established that, below the critical drive strength, the steady state of the master equation~\eqref{eqn:lindblad} in the thermodynamic limit of $\kappa \to 0^+$ can be approximated  by the steady state of the classical rate equation~\eqref{eqn:rate}.
The latter problem can be solved efficiently by replacing the exact transition rate matrix with its asymptotic form~\eqref{eqn:rate-matrix-form}, which features quasi-local hopping with asymmetry controlled by the distance to the critical point. The solution obtained in this manner shows universal behavior for large population that is independent of the truncation of the Fock space.

\section{\label{sec:steady-state}Steady-state Properties}

In this section, we discuss general properties of the steady-state solution found in the previous section. First, we propose a simple qualitative picture of the behavior of the steady state near the critical point. We then discuss the scaling of observables at the critical point.

\subsection{Qualitative Model}

In order to derive a simple qualitative picture of the physical processes at the transition, we consider a simplified model with asymmetric nearest-neighbor hopping on the semi-infinite lattice comprised of integers $n = 0,1,2, \dots$. This is meant to model the population transfer in the dressed Fock space induced by photon loss events. The corresponding classical rate equation, for $n \geq 1$, reads
\begin{equation}
   \label{eqn:hopping} \frac{d\rho_n}{dt} = \frac12\left( 1 + p \right) \rho_{n+1} + \frac12\left(1 - p \right) \rho_{n-1} - \rho_{n}.
\end{equation}
This represents a biased diffusion process on the dressed Fock space lattice depicted in Fig.~\ref{fig:state-lattice}: hopping to the left (decreasing $n$) is favored over hopping to the right (increasing $n$) via the asymmetry parameter $p>0$. This leads to a significant population of high-energy states in the steady state. 
We choose reflecting boundary conditions at the origin, so that only the first and third term in Eq. \eqref{eqn:hopping} remain for $n=0$ (with appropriate adjustment of the onsite term to maintain conservation of probability).

Equation \eqref{eqn:hopping} can be solved through the ansatz
\begin{align}
\rho_n(t) \propto e^{-qn -\gamma(q) t}
\end{align}
with ``momentum" $q$. We obtain the dispersion relation
\begin{equation}
-\gamma(q)  =\cosh q  -1 - p\sinh q.
\end{equation}
This admits two steady-state solutions with $\gamma = 0$. The first one, with $q=0$, is not permissible on the semi-infinite system as it is not normalizable. The second solution consists in choosing $q$ such that $\cosh q - p \sinh q - 1 =0$. For small $p\sim 0$, we have $q \sim 2p >0$, in which case the solution decays exponentially according to
\begin{equation}
  \label{eqn:rho-n}  \rho_n \sim e^{-2p n}.
\end{equation}
This model qualitatively reproduces two important features of the full solution to the classical master equation \eqref{eqn:rate}: (i) exponential decay in $n$ in the regime of large population ($n\ll N_{\rm max}$), see Fig.\  \ref{fig:steady-state-solutions}, (ii) scaling of the average number of excitations according to $\bar{n} \sim 1/p$, see Fig.\ \ref{fig:nbar-scaling}.

The toy model~\eqref{eqn:hopping} satisfies Kolmogorov's criterion for reversible dynamics~\cite{Tauber2014}.
That is, for any closed cycle of states in configuration space $\{s_1,s_2,...,s_M\}$, the product of transition rate matrix elements is the same in either sense this loop is traversed, so that 
\begin{align}
\nonumber & \Gamma_{s_1\to s_2} ...\Gamma_{s_{M-1} \to s_{M} } \Gamma_{s_M\to s_1} \\
& = \Gamma_{s_1 \to s_M } \Gamma_{s_{M}\to s_{M-1}}...\Gamma_{s_2\to s_1}.
\end{align}
A consequence of this is that detailed balance is satisfied. Hence, the steady-state distribution of this toy model represent an effective \emph{equilibrium} distribution $\rho_n \sim e^{-E_n/T}$, with $E_n  = n $ and effective temperature $T_{\rm eff}  \sim q \sim 2p$.

Given that this toy model appears to accurately reproduce qualitative features of the solution to the actual rate equation~\eqref{eqn:rate}, we may also question whether the solution to our problem is truly out-of-equilibrium or whether it too features emergent equilibrium behavior~\cite{Torre2013a,Marino2016,Marino2016a,Young2020,Foss-Feig2017}.
On the one hand, one can see that the transition rates of Eq.~\eqref{eqn:rate-matrix-form} do not satisfy the Kolmogorov criterion.
This may be checked for the simple three-cycle of adjacent neighbors, for which we find the asymmetry 
\begin{equation}
\frac{\Gamma_{n+1 \to n} \Gamma_{n \to n-1} \Gamma_{n-1 \to n+1}}{ \Gamma_{n-1\to n}\Gamma_{n \to n+1 }\Gamma_{n+1 \to n-1} } = \frac{(1+p)(1+p)(1-2^{\frac13} p)}{(1-p)(1-p)(1+2^{\frac13} p)} .
\end{equation}
Thus, Kolmogorov's criterion is violated and detailed balance is not satisfied for every finite $p>0$.

Nevertheless, one can verify that the steady-state distribution approximately satisfies detailed balance for $p\to 0$, at least in the region where its support is largest. Thus, it can be approximated by a ``thermal" distribution with effective temperature $T_{\rm eff} \sim \bar{n}\sim 1/p$ in the region of small $n$ (see Fig.~\ref{fig:steady-state-solutions} and Fig.~\ref{fig:nbar-scaling}).
As a result of this, the critical point describing the photon blockade breakdown may fall into a classical universality class known from equilibrium phase transitions \cite{Foss-Feig2017,Young2020}.
We leave exploring this exciting direction for future work.

\subsection{Scaling of Observables}

Using the steady-state density matrix computed in Section \ref{sec:rateeqn}, we now compute the scaling of several experimentally accessible observables in the regime  $\epsilon < \epsilon_{\rm c}$. We begin by showing that the order parameter $\langle \hat{a} \rangle $ is exactly zero in the thermodynamic limit $\kappa \rightarrow 0^+$.
This is due to the steady state respecting the $\mathbb{Z}_2$ symmetry generated by $\mathscr{C}$ (which swaps the two excitation pathways $\pm$ with each other). 
For this, it is important that the eigenstates $|\nu\rangle$ are related to the ``Fock states" $|n,s\rangle$ [Eq.\ \eqref{eqn:JCeigen}] by bosonic squeezing operators, so that  
\begin{equation}
    |\nu\rangle = \mathcal{S}(\eta) |n,s\rangle.
\end{equation}
The action of the unitary $\mathcal{S}(\eta)$ is to perform the Bogoliubov rotation so that 
\begin{equation}
    \mathcal{S}^\dagger(\eta) \hat{a} \mathcal{S}(\eta) = \cosh \eta \hat{a} + \sinh \eta \hat{a}^\dagger.
\end{equation}
Consequently, the physical photon operator obeys (see App.~\ref{app:mat-el})
\begin{equation}
\langle \nu | \hat{a} | \nu \rangle \propto \nu = s \sqrt{n}.  
\end{equation}
Thus, if $\rho_{\textrm{ss}}$ is an equal admixture of the two branches $s = \pm$ according to Eq.\ \eqref{eqn:steady-state-branch}, then the two contributions cancel, implying $ \langle \hat{a} \rangle = 0$.

We next consider the photon number, which characterizes the fluctuations of the order parameter in the disordered phase. From Eq.\ \eqref{eqn:eigen-expectations}, we obtain the expectation value in the eigenstate $|\nu\rangle$ as
\begin{equation}
    \langle \nu | \hat{a}^\dagger \hat{a} | \nu\rangle  =\left(2 \epsilon^2 + \frac12 \right)e^{2\eta} n + \frac12 e^{-2\eta} n -\frac12 \sim \frac{n}{\sqrt{1-\epsilon^2} },
\end{equation}
which is independent of the sign $s$. This implies
\begin{align}
  \nonumber  \langle \hat{a} ^\dagger \hat{a} \rangle &\propto \sum_{\nu}\langle \nu | \hat{a}^\dagger \hat{a}|\nu\rangle \rho(n)\\
    & \sim \sum_{n} \frac{n\rho (n)}{\sqrt{1-\epsilon^2}} = \frac{\bar{n}}{\sqrt{1-\epsilon^2}} \sim \frac{1}{1-\epsilon^2}.
\end{align}
The exponent which governs the divergence of the photon number upon approaching the critical point is known as the ``photon flux exponent."
We find that this exponent is 1, which is consistent with the ``classical" equilibrium Dicke model~\cite{Torre2013a}.
In this case, this is not entirely an obvious result as it arises from a combination of two distinct contributions: a power of $\frac12$ comes from the divergent number of photons in each individual eigenstate $|\nu\rangle$ and another power of $\frac12$ comes from the divergent width of the distribution $\rho(n)$ over the eigenstates. 
These contributions combine to produce the overall exponent of one. 

Finally, let us comment on the behavior of the atomic Bloch vector. 
In the steady state, we have the exact relation
\begin{equation}
\label{eqn:photon-ward-identity}
\langle [\hat{a},\hat{H} ]\rangle = -i \kappa \langle [\hat{a}^\dagger, \hat{a}] \hat{a}  \rangle . 
\end{equation}
This implies $i\kappa \langle \hat{a} \rangle = \mathcal{E} + g \langle \hat{\sigma}_-\rangle$. Since $\langle \hat{a}\rangle =0$ for $\epsilon<\epsilon_{\rm c}$, we arrive at
\begin{align}
  \label{eqn:ward}  \langle \hat{\sigma}_-\rangle = -\frac12 \epsilon.
\end{align}
The underlying physical picture is that of the Bloch vector coherently canceling out the driving field. This relation is satisfied by our mean-field theory, see Eq.\ \eqref{eqn:sigma-minus}. We can also check that the explicit steady-state solution to the rate equation \eqref{eqn:rate} satisfies this condition. For this, we use the results of App.~\ref{app:mat-el} to deduce $\langle \nu| \hat{\sigma}_x |\nu \rangle = -\epsilon$ and $\langle \nu| \hat{\sigma}_y |\nu \rangle = 0$
for all the eigenstates. 
Since the steady state is diagonal in the energy eigenbasis, we conclude that 
\begin{align}
    \langle \hat{\sigma}_x\rangle &= -\epsilon,\\
    \langle \hat{\sigma}_y\rangle &=0
\end{align}
for the steady state as well, which eventually yields Eq.\ \eqref{eqn:ward}.

Computing the expectation value $\langle \hat{\sigma}_z\rangle$ is slightly more subtle. We have $\langle \nu|\hat{\sigma}_z|\nu\rangle=0$ for every $\nu\neq 0$ . 
Thus, the only possible contribution to the steady-state expectation value comes from the dressed vacuum, which has $\nu = 0$ and produces 
\begin{align}
\langle 0 | \hat{\sigma}_z | 0 \rangle = -\sqrt{1- \epsilon^2}.
\end{align}
Thus, the steady-state expectation value of $\hat{\sigma}_z$ depends on the population of the dressed vacuum state according to
\begin{equation}
    \langle \hat{\sigma}_z \rangle = - \rho(0) \sqrt{1-\epsilon^2}.
\end{equation}
The value of $\rho(0)$ is not universal and depends on the particular approximation scheme used to compute the steady state.
However, if we model the steady-state decay in $n$ as exponential over a length $\xi\sim \bar{n}$ (as confirmed by Fig.~\ref{fig:nbar-scaling}), we can estimate that, as $\epsilon^2\rightarrow 1$, the population of the state $|0\rangle$ behaves as
\begin{equation}
\rho(0) \sim \frac{1}{\xi} \sim \frac{1}{\bar{n}}.
\end{equation}
For our steady-state solution from the rate equation we have $1/\bar{n}\sim p \sim \sqrt{1-\epsilon^2}$ and so we arrive at
\begin{equation}
 \label{eqn:sigma-z}  \langle \hat{\sigma}_z \rangle \sim  1-\epsilon^2.
\end{equation}
This is different from the mean-field theory prediction $\langle \hat{\sigma}_z \rangle_{\rm MF} \sim \sqrt{1-\epsilon^2}$. 
Although obtained by directly numerically solving the rate equation, the scaling $\bar{n} \propto 1/p$ is supported on more general grounds by the solution to the qualitative tight-binding model, the solution of which behaves as Eq.~\eqref{eqn:rho-n}.

This departure from mean-field theory is even more evident if we compute the length of the Bloch vector, $\ell^2 = |\langle \hat{\vec{\sigma}} \rangle|^2$, which also serves as a measure of the purity of the spin (upon tracing out the photon). 
Recall that in mean-field theory this is a constant of motion and so does not scale with the control parameter $\epsilon$.
Using the above results, we find instead that to leading order 
\begin{equation}
    |\langle \hat{\vec{\sigma}} \rangle|^2  \sim 1 + \mathcal{O}\left( 1-\epsilon^2\right) .
\end{equation}
Thus, the rate equation predicts that, as the critical point is approached, the length of the atomic Bloch vector approaches unity.
This implies that, as we approach the critical point, the reduced density matrix for the atom tends towards a pure state.
Knowing this may be useful for future investigations, as it motivates a natural simplifying ansatz for studying the critical behavior.

\section{\label{sec:outlook}Experimental Platforms and Outlook}

In this paper, we have provided a comprehensive analysis of the critical behavior of the driven-dissipative Jaynes--Cummings model upon approaching the breakdown of photon blockade. In the following, we discuss potential experimental platforms and point out future directions for extending our considerations.

Several experimental platforms qualify to observe this critical behavior. 
The necessary experimental ingredients are (i) a qubit degree of freedom with long coherence time, (ii) a boson with long coherence time that can be coherently driven, and (iii) the ability to realize strong coupling between the qubit and boson.

Arguably the most direct route to realize the model, as originally envisioned in Ref.~\cite{Jaynes1963}, is by using an atom, a molecule, or a quantum dot coupled to a single-mode high-quality optical cavity. Equivalently, one could couple a superconducting qubit to a microwave cavity. 
Indeed, photon blockade has been demonstrated with atoms~\cite{Birnbaum2005}, quantum dots \cite{Faraon2008}, and superconducting qubits \cite{Lang2011,hoffman11}. However, a number of obstacles remain in order to observe the critical breakdown of photon blockade in these systems, including limitations on the lifetimes of both cavity and qubit, presence of additional cavity modes, and deviations from the two-level approximation for the qubit.

Another promising candidate is to realize the system using an internal-state qubit of a trapped atomic or molecular ion coupled to its motion. If multiple ions are trapped, the qubit can be coupled to a phonon mode of the corresponding Wigner crystal.
Trapped-ion-based platforms offer a number of technical advantages, making such platforms well-suited for quantum simulation~\cite{Toyoda2013,Ivanov2009,Ge2019}, and the analogue of photon-blockade has been demonstrated in trapped ions~\cite{Debnath2018}.
Since the boson in this system is the phonon mode of the ion or of the ion crystal, the diverging boson number associated with the photon blockade breakdown may actually entail loss of the the ion from the trap or destruction of the ion crystal, though it may be possible to still get to large phonon occupation number before this occurs.

A third type of system which may be promising centers around replacing the photon (i.e.\ the bosonic subsystem) with motional quanta of a nanomechanical resonator~\cite{Rabl2009,Wallquist2010,Arcizet2011,Gieseler2019,Huillery2020}.
In order to realize the qubit, one may envision strong coupling of an atom to the nanomechanical resonator, as in Ref.~\cite{Wallquist2010}; however, a particularly promising emerging qubit candidate is the solid-state spin qubit~\cite{Rabl2009}. 
In particular, NV-centers in diamond can provide qubits with extremely long coherence times, and coupling the qubit to the nanomechanical motion has been demonstrated~\cite{Arcizet2011}.
Of special note is the coupling of the spin qubit to the motion of a levitated micromagnet~\cite{Huillery2020,Gieseler2019}, which can potentially reach the required strong-coupling regimes, while remaining well-isolated from the environment. 

Another, more abstract proposal might capitalize on the analogy between the Jaynes-Cummings model and Landau levels in graphene~\cite{Gutierrez-Jauregui2018a,Dora2009} in order to realize the critical theory in a purely solid state setting.
In this case, the bosonic degree of freedom is realized by the cyclotron motion of the electrons, while the atomic pseudo-spin degree of freedom is realized by the electron's Bloch-band degree of freedom.

In order to compare our findings to potential experiments, it is crucial to understand how deviations from the model affect the physics. In the following, we discuss several potential extensions of the model that comprise important immediate directions for future research.

First, the critical point identified in Ref. \cite{Carmichael2015} and studied here focuses on the specific point in parameter space where the bare frequencies of atom, cavity, and drive are all resonant.
This is motivated by the observation that the $\mathbb{Z}_2$ particle-hole symmetry induced by $\mathscr{C}$ is no longer a symmetry of the model when the cavity frequency is changed, while keeping the drive resonant with the atom. The rotating-frame Hamiltonian is then modified to include the cavity detuning $\delta_{\rm ph}$: 
\begin{align}
\hat{H}_{\textrm{rf}} = \delta_{\rm ph} \hat{a}^\dagger \hat{a} + g\left( \hat{a}\hat{\sigma}_{+} + \hat{a}^\dagger \hat{\sigma}_{-} \right) + \mathcal{E} \left(\hat{a}+ \hat{a}^\dagger\right).
\end{align}
While the last two terms, which are discussed in this work, are odd under $\mathscr{C}$, the first term is even. 
Thus, the $\mathbb{Z}_2$ symmetry is explicitly broken by the detuning $\delta_{\rm ph}\neq 0$, which inhibits critical fluctuations. The photon blockade breakdown transition is first-order in this case, see Refs. \cite{Dombi2015,Carmichael2015,Foss-Feig2017,Mavrogordatos2016,Young2020}.

Quite remarkably, a detuning $\delta_{\rm a}\neq 0$ of the atomic frequency with respect to the drive (while keeping the cavity resonant with the drive) does not necessarily imply a first-order transition. As was shown in Ref.~\cite{Alsing1992}, adding a finite atomic detuning still preserves the exact solubility of the driven model.
In such a case, the dressed spectrum is given by 
\begin{align}
E_{\nu} = \pm n \sqrt{1-\epsilon^2} \left[ (\delta_a/2)^2 + g^2 \sqrt{1-\epsilon^2} \right]^{1/2}. 
\end{align}
We see that, even though the $\mathscr{C}$-invariance seems to be spoiled by the detuning, there may still be a critical point due to the collapse of the eigenvalue spacing, which vanishes as $\sim \sqrt{1-\epsilon^2}$.

In our analysis, although we accounted for photon loss, we neglected spontaneous emission from the atomic degree of freedom. The inclusion of atomic decay is expected to drastically modify the critical behavior~\cite{Carmichael2015}. To understand this, recall that photon annihilation operator $\hat{a}$ essentially has no interbranch matrix elements. However, this is not true for the atomic operator $\hat{\sigma}_-$. For instance, in the absence of driving, the amplitude for switching branches due to atomic emission is $\langle n-1, - | \hat{\sigma}_- | n, +\rangle  = \frac12$,
irrespective of the value of $n$.
Contrast this to the interbranch transition amplitude for photonic emission given by $\langle n-1, - | \hat{a} | n, +\rangle = \frac{1}{2}(\sqrt{n} - \sqrt{n-1}) \sim n^{-1/2}$, Whereas this goes to zero for large quantum numbers $n$, thereby effectively decoupling the two branches when the dissipation is due to photon loss, the inclusion of atomic dissipation may qualitatively alter the steady-state properties.

Two more important effects to be considered, especially if a connection to experiment is to be made, are counter-rotating terms and the presence of multiple atomic levels.
The former has been discussed to some degree in Refs. \cite{Gutierrez-Jauregui2018,Gutierrez-Jauregui2018a}, and there is reason to believe it is not a fundamental impediment towards realizing this critical point in experiment. 
The latter may present a more important issue since many proposed qubit systems do not have the sufficiently strong non-linearity needed to project out higher internal excitations.
This is particularly troublesome in superconducting qubits~\cite{Bishop2009,Fink2017,Raftery2014,Lang2011,Fitzpatrick2017,Vaneph2018}, where higher lying levels can present complications to the effective Jaynes-Cummings picture.

Even within the confines of the present model, several intriguing and experimentally relevant open problems  remain.
For instance, how does the transition look when approached from the ``other side", i.e.\ for $\epsilon^2 > 1$. 
Based on our semiclassical calculations, as well as our argument based on the rate equation, we should not expect this part of the phase diagram to be well defined if we insist on taking the $\kappa\to 0^+$ limit.
Nevertheless, we can still study the behavior of the steady-state for finite $\kappa$.
In this case, we expect the critical point to evolve into a smooth crossover.
Studying this in a quantitative manner may be achieved in terms of a functional integral description~\cite{Kamenev2011,Diehl16,Foss-Feig2017,Young2020,Torre2013a}, by building upon the saddle-point semiclassical solutions, now including fluctuations due to finite $\kappa$.
This has the additional advantage in that it allows us to study dynamical correlation functions whereas in this work we are limited to only studying static correlation functions. 
This may be particularly relevant given that the steady state appears to admit an emergent equilibrium description~\cite{Foss-Feig2017,Young2020}, despite the microscopic violation of detailed balance.
Studying the dynamical correlations of the system can then allow for determining whether the system obeys a generalized fluctuation-dissipation relation, as is often the case when systems display emergent equilibrium~\cite{Foss-Feig2017,Diehl16,Torre2013a,Paz,Tauber2014}.

This approach may also naturally lend itself towards a generalization to include a many-body version of this transition.
Specifically, we might imagine a Bose--Hubbard-like system where, instead of an onsite Hubbard non-linearity of the form $\hat{a}^\dagger \hat{a}^\dagger \hat{a}\hat{a}$, we consider an onsite Jaynes--Cummings-type non-linearity. 
Such a model has been considered in Refs.~\cite{Koch2009,Zhu2013,Schmidt2013,Schmidt2010,Minar2016}, but not in the context of the breakdown of photon blockade.
In principle, this could dramatically alter the ground-state phase diagram. 
To see why, we might imagine a quantum fluctuation in the order parameter of a neighboring site which is sufficiently strong as to induce a breakdown of the photon blockade.
This would result in a destruction of the Mott state locally.
In principle, these tunneling events may then proliferate and ultimately destabilize the Mott lobe entirely.

\begin{acknowledgments}
We thank Oles Shtanko, Victor Galitski, Daniel Paz, Colin Rylands, Tibor Rakovsky, Rex Lundgren, Mohammad Hafezi, and Christopher Flower for fruitful discussions. 
I.B., J.T.Y., and A.V.G.\ acknowledge funding by AFOSR MURI, AFOSR, NSF PFCQC program, DoE BES Materials and Chemical Sciences Research for Quantum Information Science program (award No.\ DE-SC0019449), ARO MURI, ARL CDQI, DoE ASCR Quantum Testbed Pathfinder program (award No.\ DE-SC0019040), DoE ASCR FAR-QC program (award No.\ DE-SC0020312), and NSF PFC at JQI.
J.B.C.\ received support from the U. S. Army  Research Laboratory and the U. S. Army Research Office under contract number W911NF1810164, NSF DMR-1613029, and the Simons Foundation.
Part of this work was performed at the Kavli Institute for Theoretical Physics with support from the Heising-Simons Foundation, the Simons Foundation, and National Science Foundation Grant No. NSF PHY-1748958.
H.J.C.\ acknowledges the support of the New Zealand Tertiary Education Committee through the Dodd-Walls Centre for Photonic and Quantum Technologies. M.M. acknowledges support from NSF under Grant No. DMR-1912799,  the Air Force Office of Scientific Research (AFOSR) under award number FA9550-20-1-0073 as well as the start-up funding from Michigan State University.

\end{acknowledgments}

\appendix
\section{\label{app:drive-eigen}Driven Eigenstates}

We follow Ref.~\cite{Alsing1992} and obtain the eigenstates and eigenvalues of the Hermitian operator (rescaled by $g$)
\begin{equation} 
\hat{H} = \hat{a}^\dagger \hat{\sigma}_{-} + \hat{a}\hat{\sigma}_+ + \frac12 \epsilon(\hat{a}+\hat{a}^\dagger)
\end{equation}
with 
\begin{equation}
    \epsilon \equiv 2\mathcal{E}/g
\end{equation}
the unitless control parameter. 
We will henceforth, without loss of generality, consider $\epsilon > 0$.
We use the convention for the displacement and squeezing operators ($\eta \in\mathbb{R}$)
\begin{equation} 
\begin{aligned}
& \mathcal{D}(\alpha) = \exp\left( \alpha \hat{a}^\dagger - \alpha^* \hat{a}^\dagger \right),\\
& \mathcal{S}(\eta) =  \exp\left( \frac12 \eta (\hat{a}^\dagger \hat{a}^\dagger - \hat{a}\hat{a}  )  \right), \\
\end{aligned}
\end{equation}
which act on the annihilation operator as  
\begin{align}
& \mathcal{D}^\dagger(\alpha) \hat{a} \mathcal{D}(\alpha) = \hat{a} + \alpha,  \\
& \mathcal{S}^\dagger(\eta) \hat{a} \mathcal{S}(\eta) = u\hat{a} + v \hat{a}^\dagger,  
\end{align}
with $ u = \cosh(\eta)$ and $v = \sinh(\eta)$.

Let us consider the eigenvalue problem 
\begin{equation}
\left( \hat{a}^\dagger \hat{\sigma}_{-} + \hat{a}\hat{\sigma}_+  + \frac12 \epsilon (\hat{a}+\hat{a}^\dagger) - \lambda_\nu \right) |\nu \rangle = 0.
\end{equation}
We will use the convention that $\nu$ labels the eigenstates and $\lambda_\nu$ is the eigenvalue for eigenstate $\nu$.

First, with the knowledge of the exact solution, we perform a squeezing transformation of the form 
\begin{equation}
|\nu \rangle \equiv \mathcal{S}(\eta)|n,s\rangle .
\end{equation}
We will later determine the exact squeezing parameter $\eta$.
For the moment, the notation $|n,s\rangle$ is merely suggestive, but, as we will see, this is consistent with the quantum number labeling scheme used throughout the paper. 
This transformation acts on $\hat{H}$ to produce 
\begin{equation}
\left( \hat{\tau}_{+} \hat{a} + \hat{\tau}_{-} \hat{a}^\dagger  - \lambda \right) |n,s\rangle = 0,
\end{equation} 
where we have defined the new two-by-two matrix
\begin{equation} 
\hat{\tau}_{+} = \hat{\tau}_{-}^\dagger \equiv u \hat{\sigma}_{+}+ v \hat{\sigma}_{-} + \frac12\epsilon(u+v) .
\end{equation}
We observe that 
\begin{equation} 
\det\hat{\tau}_{-} = \frac14 \epsilon^2 (u^2 + v^2) -(1-\frac12\epsilon^2) uv,
\end{equation}
hence, the matrix $\hat{\tau}_{-}$ is singular when  
\begin{equation}
\det\hat{\tau}_{-} = 0 \Leftrightarrow \tanh(2\eta) = \frac{\epsilon^2}{2 - \epsilon^2}.
\end{equation}
This only has a solution if $0 < \epsilon^2 /(2-\epsilon^2) < 1 \Rightarrow \epsilon^2 < 1$. 
If this condition is met, then we can choose $\eta$ such that 
\begin{equation} 
\left(\frac12 \epsilon (u+v)\right)^2 = uv \Rightarrow \frac12 \epsilon (u+v) = \sqrt{uv}.
\end{equation}
This leads to the following matrix representation for $\hat{\tau}_{-}$:  
\begin{align}
\hat{\tau}_{-} = \left(\begin{array}{cc}
\sqrt{uv} 	&	v	\\
u	&	\sqrt{uv} 	\\
\end{array}\right).
\end{align}
Let $\chi_R, \chi_L^\dagger$ be the normalized left- and right-eigenvectors with zero eigenvalue respectively, so that 
\begin{align}
\hat{\tau}_{-}\chi_R = \chi_L^\dagger \hat{\tau}_{-} = 0 .
\end{align}
Note that, by virtue of $\hat{\sigma}_x \hat{\tau}_{+}  \hat{\sigma}_x = \hat{\tau}_{-}$, these are related by $\chi_L = \hat{\sigma}_x \chi_R$.
We have the specific representation of $\chi_R$ as 
\begin{equation} 
\chi_R = \left(\begin{array}{c}
-\sqrt{\frac{v}{u+v}} \\
\sqrt{\frac{u}{u+v} } \\
\end{array}\right).
\end{equation}

We begin by determining the ground-state (vacuum) $|0\rangle$, as the state which is annihilated by both $\hat{\tau}_{-}$ and $\hat{a}$. 
The eigenvalue is zero and the ket is
\begin{equation} 
|0 \rangle = \chi_R \otimes |\textrm{vac}\rangle,  
\end{equation}
where $|\textrm{vac}\rangle$ is the photon vacuum state.

Next we determine the excitations above the vacuum. 
These are organized into two-dimensional sub-spaces of the form
\begin{equation} 
|n,s\rangle = A  \chi_R \otimes |\psi^R\rangle + B  \chi_L\otimes |\psi^L\rangle.
\end{equation}
We insert this ansatz and then act on the left with $\chi_R^\dagger$ and $\chi_L^\dagger$.
Using that these are zero left-eigenvectors of $\hat{\tau}_{+}$ and $\hat{\tau}_{-}$ respectively, we find 
\begin{equation}
\begin{aligned}
& B  \chi_R^\dagger \hat{\tau}_{-}\chi_L \left(\hat{a}^\dagger|\psi^L\rangle\right) = \lambda \left( A |\psi^R \rangle + B \chi_R^\dagger \chi_L|\psi^L\rangle \right), \\
& A \chi_L^\dagger \hat{\tau}_{+}\chi_R \left(\hat{a}|\psi^R\rangle\right) = \lambda \left( A\chi_L^\dagger\chi_R |\psi^R \rangle + B |\psi^L\rangle \right). \\
\end{aligned}
\end{equation}
These may be decoupled to obtain the equation for the ``left" ket as  
\begin{equation}
\label{eqn:left-ket}
|\psi^L\rangle = \frac{A}{B}\left( \frac{1}{\lambda}\chi_L^\dagger \hat{\tau}_{+} \chi_R \hat{a} - \chi_L^\dagger \chi_R\right) |\psi^R\rangle 
\end{equation}
and the equation for the ``right" ket as 
\begin{equation}
\label{eqn:right-ket}
\begin{aligned}
& \left( \chi_R^\dagger \hat{\tau}_{-}\chi_L \hat{a}^\dagger- \lambda \chi_R^\dagger\chi_L \right)\left(\chi_L^\dagger\hat{\tau}_{+}\chi_R \hat{a} -\lambda \chi_L^\dagger \chi_R \right) |\psi^R \rangle \\
& = \lambda^2 |\psi^R\rangle .\\
\end{aligned}
\end{equation}
We can solve this by a displaced Fock state such that 
\begin{equation}
|\psi^R\rangle = \mathcal{D}\left( \alpha\right)|n\rangle, 
\end{equation}
with 
\begin{equation}
    \alpha = \lambda \frac{\chi_L^\dagger \chi_R }{|\chi_L^\dagger \hat{\tau}_{+} \chi_R |}
\end{equation}
the displacement and 
\begin{equation}
    \lambda^2 = n |\chi_L^\dagger \hat{\tau}_{+}\chi_R|^2 
\end{equation}
the eigenvalue, with $n = 0,1,2,3...$. 
This has two branches of solution, 
\begin{equation} 
\lambda = \pm \sqrt{n} | \chi_R^\dagger \hat{\tau}_{-} \chi_L |.
\end{equation}
The kets $|n,s\rangle$ are then determined by using equation~\eqref{eqn:left-ket} and normalizing.

We thus obtain the final result for the spectrum as 
\begin{equation}
|\nu \rangle =  \mathcal{S}(\eta) |n,s\rangle,  
\end{equation}
with ``ladder" states 
\begin{equation} 
\label{eqn:ladder-states}
\begin{aligned}
& |n,s\rangle =\mathcal{D}(\alpha_{\nu}) \frac{1}{\sqrt{2}} \left( \chi_R |n\rangle +s \chi_L |n-1\rangle \right), \\
& |0\rangle = \chi_R  | \textrm{vac} \rangle,  \\
\end{aligned}
\end{equation}
where $s=\pm $ and $ n = 1,2,....$.
The squeezing parameter is obtained as
\begin{equation} 
\eta = -\frac14 \log (1-\epsilon^2), 
\end{equation}
and $\alpha_{\nu},\lambda_\nu$ are 
\begin{equation}
\begin{aligned}
& \lambda_\nu =  s\sqrt{n} e^{-3\eta}  = s \sqrt{n}\left(1-\epsilon^2\right)^{\frac34} = \nu \left(1-\epsilon^2\right)^{\frac34},  \\
& \alpha_{\nu} = -\epsilon e^{3\eta }\lambda = -\epsilon s\sqrt{n} = -\epsilon \nu,\\
\end{aligned}
\end{equation}
and, as usual, $u=\cosh\eta,v=\sinh\eta$.

\section{\label{app:mat-el}Matrix Elements}

In this Appendix, we compute the matrix elements of various operators in the driven-eigenbasis computed in Sec.~\ref{app:drive-eigen}. 

Let us start with the ``vacuum state" with $\nu = 0$ as it is simpler and somewhat of a special case.
For the expectation values of $\hat{a}$ and $\hat{a}^\dagger \hat{a}$, we have 
\begin{equation}
\langle 0 | \hat{a} | 0 \rangle = \langle \textrm{vac} | u\hat{a}+v\hat{a}^\dagger |  \textrm{vac} \rangle = 0 
\end{equation}
and 
\begin{equation}
\langle 0 | \hat{a}^\dagger \hat{a} |0 \rangle =\langle\textrm{vac} | \left( u\hat{a}^\dagger+v\hat{a}\right)\left( u\hat{a}+v\hat{a}^\dagger\right) |  \textrm{vac} \rangle = v^2 .
\end{equation}
Recalling that $v = \sinh \eta$, we obtain the scaling upon approaching the critical point of 
\begin{equation}
\langle 0 |\hat{a}^\dagger \hat{a} |0 \rangle \sim \frac{1}{\sqrt{1-\epsilon^2}}.
\end{equation}
For the expectation values of the spin operators, we have a relatively simple calculation of 
\begin{equation}
    \langle 0 | \hat{\vec{\sigma}} | 0 \rangle = \chi_R^\dagger  \hat{\vec{\sigma}} \chi_R = \left(-\epsilon, 0 , -\sqrt{1-\epsilon^2} \right) .
\end{equation}

For the states with $\nu \neq 0$, the result is a bit more complicated. 
We first perform the squeezing transformation to go from the eigenstates $|\nu\rangle$ to the ladder states $|n,s\rangle $ to obtain
\begin{align}
\langle \nu | \hat{a} | \nu\rangle &= \langle n,s | u \hat{a} + v\hat{a}^\dagger | n,s\rangle,  \\
\langle \nu | \hat{a}^\dagger \hat{a} | \nu\rangle &= \langle n,s | \left(u \hat{a} + v\hat{a}^\dagger\right)\left( u \hat{a} + v\hat{a}^\dagger\right) | n,s\rangle,  \\
\langle \nu | \hat{\vec{\sigma}} | \nu\rangle &= \langle n,s | \hat{\vec{\sigma}} | n,s\rangle .
\end{align}

The field expectation value is the simplest, and can be computed to be
\begin{equation}
    \langle n,s | \hat{a} |n,s \rangle = \frac32 \alpha_{\nu}.
\end{equation}
Similarly, the spin operators are relatively simple to calculate and yield 
\begin{equation}
    \langle n,s | \hat{\vec{\sigma}}  |n,s \rangle = \left( - \epsilon , 0 ,0 \right).
\end{equation}

For the photon number, due to the Bogoliubov transformation, we require both of the expectation values
\begin{align}
\langle n ,s | \{\hat{a}^\dagger, \hat{a}\} |n,s\rangle &= 2n + 4\alpha_{\nu}^2, \\
\langle n ,s | \hat{a}^2 |n,s\rangle &=  2\alpha_{\nu}^2.
\end{align}
We combine these with the behavior of the Bogoliubov coefficients $ u =\cosh \eta$ and $v = \sinh \eta$ to obtain 
\begin{align}
\label{eqn:eigen-expectations}
\langle \nu | \hat{a} | \nu\rangle &= \frac32 \alpha_\nu e^{\eta}, \\
\langle \nu | \hat{a}^\dagger \hat{a} | \nu\rangle &= \left(\frac{n}{2} + 2\alpha^2\right)e^{2\eta} - \frac12 + \frac{n}{2}e^{-2\eta},    \\
\langle \nu | \hat{\vec{\sigma}} | \nu\rangle &= \left( - \epsilon , 0 ,0 \right).
\end{align}
The results for both the $\nu = 0$ and $\nu \neq 0 $ states are summarized in Tab.~\ref{tab:expectations}. 

\renewcommand{\arraystretch}{1.5}
\begin{table}[t!]
    \begin{tabular}{| c | c |}
    \hline
     $\nu = 0$ &  $\nu \neq 0$  \\
    \hline
    \hline
    $\langle 0 | \hat{a} | 0 \rangle =0  $ & $\langle \nu | \hat{a} | \nu \rangle = -\frac32\epsilon e^{\eta} \nu$  \\
    $\langle 0 | \hat{a}^\dagger \hat{a} | 0 \rangle = \sinh^2 \eta $ & $ \langle \nu | \hat{a}^\dagger \hat{a} | \nu \rangle =  \left( 2\epsilon^2e^{2\eta} +\cosh 2\eta \right) \nu^2 - \frac12 $ \\
    $\langle 0 | \hat{\sigma}_x | 0 \rangle= -\epsilon $  & $\langle \nu | \hat{\sigma}_x | \nu \rangle= -\epsilon $  \\
    $\langle 0 | \hat{\sigma}_y | 0 \rangle= 0  $  & $\langle \nu |  \hat{\sigma}_y | \nu \rangle= 0 $  \\
    $\langle 0 | \hat{\sigma}_z | 0 \rangle=  -\sqrt{1-\epsilon^2} $  & $\langle \nu | \hat{\sigma}_z | \nu \rangle = 0 $  \\
    \hline
    \end{tabular}
    \caption{Expectation values of different observables in the dressed eigenstates. }
    \label{tab:expectations}
\end{table}
\renewcommand{\arraystretch}{1}

Next, we compute the transition elements between eigenstates via single-photon emission events.  
Let $\nu$ and $\mu$ be the two eigenvalues connected by the transition. 
Then we need the off-diagonal elements for the transition $\mu \rightarrow \nu$.
These are expressed in terms of the ladder states of Eq.~\eqref{eqn:ladder-states} as $\langle \nu| \hat{a} | \mu \rangle = \langle n,s| u\hat{a} + v\hat{a}^\dagger |m,r\rangle$, where $\nu = s\sqrt{n}$ is the final quantum number and $\mu = r\sqrt{m}$ is the initial quantum number.
To compute $\langle \nu| \hat{a} | \mu \rangle$, define the ladder-state matrix elements of $\hat{a}$ as
\begin{equation}
    A_{n,s|m,r} = \langle n,s| \hat{a} | m,r\rangle.
\end{equation}
This quantity is regular at the transition point and so we set $\epsilon = 1$. We have
\begin{widetext}
\begin{align}
 \nonumber A_{n,s|m,r} &= \frac{1}{2} \Bigl(\langle n|-s\langle n-1|\Bigr)\mathcal{D}(-s\sqrt{n})^\dagger \hat{a}  \mathcal{D}(-r\sqrt{m})\Bigl(|m\rangle - r |m-1\rangle\Bigr)\\
 &=  \frac{1}{2} \Bigl(\langle n|-s\langle n-1|\Bigr)\mathcal{D}(-s\sqrt{n})^\dagger  \mathcal{D}(-r\sqrt{m})(\hat{a}-r\sqrt{m})\Bigl(|m\rangle - r |m-1\rangle\Bigr).
\end{align} 
Due to orthogonality
\begin{align}
 \Bigl(\langle n|-s\langle n-1|\Bigr)\mathcal{D}(-s\sqrt{n})^\dagger \mathcal{D}(-r\sqrt{m})\Bigl(|m\rangle - r |m-1\rangle\Bigr)=0,
\end{align} 
this simplifies to
\begin{align}
 \nonumber A_{n,s|m,r}   &= \frac{1}{2} \Bigl(\langle n|-s\langle n-1|\Bigr)\mathcal{D}(\alpha)  \Bigl(\sqrt{m}|m-1\rangle - r\sqrt{m-1} |m-2\rangle\Bigr)\\
  &= \frac{1}{2} \langle 0| \Bigl(\frac{1}{\sqrt{n!}}\hat{a}^n - s\frac{1}{\sqrt{(n-1)!}}\hat{a}^{n-1}\Bigr)\mathcal{D}(\alpha)  \Bigl(\frac{\sqrt{m}}{\sqrt{(m-1)!}}(\hat{a}^\dagger)^{m-1} - r\frac{\sqrt{m-1}}{\sqrt{(m-2)!}} (\hat{a}^\dagger)^{m-2} \Bigr)|0\rangle,
\end{align}
with 
\begin{equation}
    \alpha=s\sqrt{n}-r\sqrt{m}.
\end{equation}
\end{widetext}

In order to evaluate this expression, we employ the generating function
\begin{equation}
\langle 0 | \hat{a}^{p}\mathcal{D}(\alpha) \hat{a}^{\dagger (q)} |0\rangle = e^{|\alpha|^2/2}\left(\frac{\partial}{\partial \alpha} \right)^{q}\left(-\frac{\partial }{\partial \alpha^*}\right)^{p} e^{-|\alpha|^2} .
\end{equation}
The derivative with respect to $\alpha^*$ may be performed to obtain
\begin{align}
\langle 0 | \hat{a}^{p}\mathcal{D}(\alpha) \hat{a}^{\dagger (q)} |0\rangle  = e^{|\alpha|^2/2}\left(\frac{\partial}{\partial \alpha} \right)^{q}\left(\alpha^{p} e^{-\alpha^* \alpha}\right).
\end{align}
As the derivative treats $\alpha$ and $\alpha^*$ independently, we can now rescale $\alpha$ by $\alpha^*$ via $\alpha = r/\alpha^*$ which then allows us to write 
\begin{multline}
\langle 0 | \hat{a}^{p}\mathcal{D}(\alpha) \hat{a}^{\dagger (q)} |0\rangle  = e^{r/2}\left(\alpha^*\right)^{q-p}\left(\frac{\partial}{\partial r} \right)^{q}\left(r^{p} e^{-r}\right) \\
= q!  e^{-r/2} (\alpha^2)^{p-q}(\alpha^*)^{q-p} L_{q}^{(p-q)}(r),
\end{multline}
with associated Laguerre polynomials $L_n^{(k)}(z)$. 
The second equality follows by application of the Rodrigues formula for associated Laguerre polynomials.
We can now take $\alpha$ to be real to obtain the result 
\begin{equation}
    \langle 0 | \hat{a}^{p}\mathcal{D}(\alpha) \hat{a}^{\dagger (q)} |0\rangle = e^{-\alpha^2/2} q! \alpha^{p-q}L_{q}^{(p-q)}(\alpha^2).
\end{equation}

We thus obtain the manifestly real expression 
\begin{align}
  \nonumber A_{n,s|m,r} ={}& \frac{e^{-\alpha^2/2}}{2} \Biggl[ \sqrt{\frac{m!}{n!}} \alpha^{n-m+1} L^{(n-m+1)}_{m-1}(\alpha^2) \\
  \nonumber &-s \sqrt{\frac{m!}{(n-1)!}} \alpha^{n-m} L^{(n-m)}_{m-1}(\alpha^2)\\
 \nonumber &-r\sqrt{\frac{(m-1)!}{n!}} \alpha^{n-m+2} L^{(n-m+2)}_{m-2}(\alpha^2)\\
 \label{eqn:EqAnm} &+rs \sqrt{\frac{(m-1)!}{(n-1)!}} \alpha^{n-m+1} L^{(n-m+1)}_{m-2}(\alpha^2)\Biggr].
\end{align}
Note that the matrix elements satisfy
\begin{equation}
A_{n,-s|m,-r} = (-1)^{n-m+1} A_{n,s|m,r},
\end{equation}
and so we only need to consider two out of four possible combinations of the signs $r$ and $s$: one same-sign combination and one opposite-sign combination.

We are interested in the asymptotic behavior of $A_{n,s|m,r}$ for large $m,n\to \infty$ as a function of the difference $m-n$. We consider the case of equal sign and set $r=s=1$. Define the relative coordinate
\begin{align}
 R= \sqrt{n}-\sqrt{m},
\end{align}
so that $\alpha=R$ for our sign choice. We have
\begin{align}
 \label{eqn:Anm} A_{n,1|m,1} &= A_{nm}^{(1)} +  A_{nm}^{(2)} + A_{nm}^{(3)} + A_{nm}^{(4)}
\end{align}
with
\begin{align}
  A_{nm}^{(1)} &=  \frac{e^{-R^2/2}}{2} \sqrt{\frac{m!}{n!}} R^{n-m+1} L^{(n-m+1)}_{m-1}(R^2),\\
 A_{nm}^{(2)} &=  -\frac{e^{-R^2/2}}{2} \sqrt{\frac{m!}{(n-1)!}} R^{n-m} L^{(n-m)}_{m-1}(R^2),\\
 A_{nm}^{(3)} &= - \frac{e^{-R^2/2}}{2} \sqrt{\frac{(m-1)!}{n!}} R^{n-m+2} L^{(n-m+2)}_{m-2}(R^2),\\  
  A_{nm}^{(4)} &=  \frac{e^{-R^2/2}}{2} \sqrt{\frac{(m-1)!}{(n-1)!}} R^{n-m+1} L^{(n-m+1)}_{m-2}(R^2).
\end{align}
We write $m=n+\delta$ with integer $\delta \neq 0$ so that
\begin{align}
 A_{n,n+\delta}^{(1)} &=  \frac{e^{-R^2/2}}{2} \sqrt{\frac{(n+\delta)!}{n!}} R^{1-\delta} L^{(1-\delta)}_{n+\delta-1}(R^2),\\
 A_{n,n+\delta}^{(2)} &=  -\frac{e^{-R^2/2}}{2} \sqrt{\frac{(n+\delta)!}{(n-1)!}} R^{-\delta} L^{(-\delta)}_{n+\delta-1}(R^2),\\
 A_{n,n+\delta}^{(3)} &= - \frac{e^{-R^2/2}}{2} \sqrt{\frac{(n+\delta-1)!}{n!}} R^{2-\delta} L^{(2-\delta)}_{n+\delta-2}(R^2),\\  
  A_{n,n+\delta}^{(4)} &= \frac{e^{-R^2/2}}{2} \sqrt{\frac{(n+\delta-1)!}{(n-1)!}} R^{1-\delta} L^{(1-\delta)}_{n+\delta-2}(R^2).
\end{align}
To study the asymptotic behavior of these expressions in the limit $n\sim \infty$ while keeping $\delta$ fixed, we use
\begin{align}
 R &= \sqrt{n} - \sqrt{n+\delta}  \sim - \frac{\delta}{2\sqrt{n}}
\end{align}
and
\begin{align}
 \sqrt{\frac{(n+c)!}{(n+d)!}} \sim \sqrt{n}^{c-d},\ e^{-R^2/2} \sim 1 - \frac{\delta^2}{8n} \sim 1.
\end{align}
We then have
\begin{align}
 A_{n,n+\delta}^{(1)} &\sim \frac{1}{2} \Bigl(- \frac{\delta}{2\sqrt{n}}\Bigr)^{1-\delta} \sqrt{n}^\delta L_{n+\delta-1}^{(1-\delta)}\Bigl(\frac{\delta^2}{4n}\Bigr).
\end{align}
The asymptotic behavior of the Laguerre polynomials $L_N^{(k)}$ for $x>0$, $N\to \infty$ and $k$ fixed is determined by
\begin{align}
 L^{(k)}_N\Bigl(\frac{x}{N}\Bigr) \sim \Bigl(\frac{N}{\sqrt{x}}\Bigr)^k e^{\frac{x}{2N}} J_k(2\sqrt{x}),
\end{align}
with $J_k(y)$ the Bessel function. Note that
\begin{align}
 J_k(-x) = (-1)^k J_k(x) = J_{-k}(x).
\end{align}
Hence
\begin{align}
 L_{n+\text{const}}^{(k)}\Bigl(\frac{\delta^2}{4n}\Bigr) &\sim \Bigl(\frac{2n}{|\delta|}\Bigr)^{k} e^{\frac{\delta^2}{8n}} J_{k}(|\delta|)\sim \Bigl(\frac{2n}{|\delta|}\Bigr)^{k} J_{k}(|\delta|),
\end{align}
and so
\begin{align}
 \nonumber A_{n,n+\delta}^{(1)} &\sim  \frac{\sqrt{n}}{2} \Bigl(-\sgn(\delta)\Bigr)^{1-\delta}J_{1-\delta}(|\delta|)\\
 \nonumber &=\frac{\sqrt{n}}{2} \Bigl(-\sgn(\delta)\Bigr)^{1-\delta}\Bigl(\sgn(\delta)\Bigr)^{1-\delta} J_{1-\delta}(\delta)\\
 \nonumber &=\frac{\sqrt{n}}{2} (-1)^{1-\delta} J_{1-\delta}(\delta)\\
  &=\frac{\sqrt{n}}{2} J_{\delta-1}(\delta).
\end{align}
Similarly, we obtain
\begin{align}
 A_{n,n+\delta}^{(2)} &\sim -\frac{\sqrt{n}}{2}  J_{\delta}(\delta), \\
 A_{n,n+\delta}^{(3)} &\sim -\frac{\sqrt{n}}{2}  J_{\delta-2}(\delta),\\  
  A_{n,n+\delta}^{(4)} &\sim \frac{\sqrt{n}}{2}  J_{\delta-1}(\delta).
\end{align}
We conclude that the asymptotic behavior of Eq.\ (\ref{eqn:Anm}) for for $n\to \infty$ with $m=n+\delta$ and $\delta\in\mathbb{Z}$ is given by
\begin{align}
 A_{n,1|n+\delta,1} \sim \sqrt{n} f(\delta),
\end{align}
where
\begin{equation}
 \label{eqn:scaling-f} f(\delta) = \frac{1}{\delta}J_{\delta-1}(\delta).
\end{equation}
The first line follows from Eq.\ \eqref{eqn:EqAnm} in the limit $\alpha\to 0$, and the second line follows from $J_{k-1}(y)+J_{k+1}(y) = \frac{2k}{y}J_k(y)$ with $k=\delta-1$.

We next justify neglecting the interbranch transitions by showing that they decay exponentially in $n$ for large $n, m$. For this purpose, we choose $s=1$ and $r=-1$ in Eq.\ \eqref{eqn:EqAnm}. Setting $n=m$ (corresponding to $\delta=0$), we have $\alpha=2\sqrt{n}$ and arrive at
\begin{align}
  \nonumber  A_{n,1|n,-1} &\sim \frac{\sqrt{n} e^{-2n}}{2} \Bigl[ 2L_{n-1}^{(1)}(4n)-L_{n-1}(4n) \\
    &+ 4 L_{n-2}^{(2)}(4n) -2 L_{n-2}^{(1)}(4n)\Bigr].
\end{align}
We numerically fit the asymptotic behavior for large $n\in[10,100]$ to be
\begin{align}
  \label{eqn:inter}  A_{n,1|n,-1} \sim (-1)^{n+1} \cdot 0.07 n^{-0.83},
\end{align}
which is an excellent approximation even for $n$ of order unity. Consequently, $|A_{n,1|n,-1}|^2$ decays faster than $1/n$ For $\delta\in\mathbb{Z}$, $\delta\neq 0$, we have $\alpha = \sqrt{n} + \sqrt{n+\delta} \sim 2\sqrt{n}+\frac{\delta}{2\sqrt{n}}$, and verify numerically that $|A_{n,1|n+\delta,-1}|$ decays with an exponent close to the one in Eq.\ \eqref{eqn:inter}.

We now further evaluate the asymptotic matrix elements for moderate to large $\delta$, which leads to a simple power-law form. For this purpose, we separate the function $f(\delta)$ in \eqref{eqn:scaling-f} into even and odd parts according to
\begin{align}
    f(\delta) = s(|\delta|) + \sgn(\delta) a(|\delta|)
\end{align}
with
\begin{align}
    s(y) &= \frac{1}{2y}\Bigl[  J_{y-1}(y) -J_{y+1}(y)\Bigr],\\
    a(y) &= \frac{1}{2y}\Bigl[ J_{y-1}(y) +J_{y+1}(y)\Bigr].
\end{align}
Note that, while $f(\delta)$ is oscillatory for real $\delta<0$, we only evaluate the function for integer arguments $\delta\in \mathbb{Z} \backslash \{0\}$, where these oscillations are not visible. For $y>0$ we use the integral representation of the Bessel functions to write
\begin{align}
    J_{y\pm 1}(y) = \int_{-\pi}^\pi \frac{\mbox{d}\theta}{2\pi} e^{-\rmi \delta [\sin(\theta)-\theta]\mp \rmi \theta}.
\end{align}
For large $\delta\to \infty$, the integral will be dominated by the region of small angles. We expand $\sin(\theta) - \theta = -\frac{1}{6}\theta^3+\mathcal{O}(\theta^5)$ and introduce the variable $x^3 = \frac{y}{2}y^3$ and extend the integration boundaries to infinity. We then arrive at
\begin{align}
  \nonumber  J_{y\pm 1}(y) &= \Bigl(\frac{2}{y}\Bigr)^{1/3}\int_{-\infty}^\infty \frac{\mbox{d}\theta}{2\pi} e^{-\rmi x^3/3\mp \rmi (\frac{2}{y})^{1/3}x}\\
  &= \Bigl(\frac{2}{y}\Bigr)^{1/3} \text{Ai}\Bigl(\mp (2/y)^{1/3}\Bigr),
\end{align}
with Airy function
\begin{align}
 \nonumber   \text{Ai}(x) &= \int_0^\infty \frac{\mbox{d}\theta}{\pi}\cos\Bigl( \frac{\theta^3}{3}+x\theta\Bigr)\\
    &=\frac{1}{3^{2/3}\Gamma(\frac{2}{3})} +  \frac{1}{3^{1/3}\Gamma(\frac{1}{3})}x+\mathcal{O}(x^2)
\end{align}
and Euler's function $\Gamma(x)$. Consequently, as $y\to \infty$, we have
\begin{align}
    s(y) &\sim \frac{(4/3)^{1/3}}{\Gamma(\frac{1}{3})} y^{-5/3},\\
    a(y) &\sim \frac{(2/9)^{1/3}}{\Gamma(\frac{2}{3})} y^{-4/3}.
\end{align}
In fact, these formulas are excellent approximations even for small $y\sim \mathcal{O}(1)$. We further have
\begin{align}
    \frac{a(y)}{s(y)} \sim \frac{\Gamma(\frac{1}{3})}{6^{1/3}\Gamma(\frac{2}{3})} y^{1/3} \approx y^{1/3},
\end{align}
where the prefactor is close to unity.

\section{Dynamic Stability of Mean-Field Equations}\label{app:stability}
In this section, we investigate the stability of the set of mean-field equations \eqref{eqn:mean-field}. We write
\begin{align}
    a & =\langle \hat{a}\rangle,\    \sigma = \langle \hat{\sigma}_-\rangle,\    \sigma_3 = \langle \hat{\sigma}_z\rangle.
\end{align}
The flow equations in these variables read
\begin{align}
 \dot{a} &= -\kappa a -\rmi(\mathcal{E}+g\sigma),\\
 \dot{\sigma} &= \rmi g a \sigma_3,\\
 \dot{\sigma}_3 &= -2\rmi g(a \sigma^*-a^*\sigma). 
\end{align}
We assume $\mathcal{E},g>0$. We then decompose $a$ and $\sigma$ into their real and imaginary parts according to
\begin{align}
 a &= a_1 +\rmi a_2,\\
 \sigma &=\sigma_1 + \rmi \sigma_2,
\end{align} 
to arrive at
\begin{align}
 \dot{a} &= -\kappa(a_1+\rmi a_2) +(-\rmi \mathcal{E} +g\sigma_2 -\rmi g \sigma_1),\\
 \dot{\sigma} &= \rmi g(a_1+\rmi a_2)\sigma_3,\\
 \dot{\sigma}_3 & = 4g (a_2\sigma_1-a_1\sigma_2) .
\end{align}
Hence the flow equations for the five real parameters $\vec{c}=(a_1,a_2,\sigma_1,\sigma_2,\sigma_3)$ read
\begin{align}
 \dot{a}_1 &= -\kappa a_1 + g\sigma_2,\\
 \dot{a}_2 &= -\kappa a_2 -\mathcal{E} -g \sigma_1,\\
 \dot{\sigma}_1 &= -g a_2\sigma_3,\\
 \dot{\sigma}_2 &= g a_1 \sigma_3,\\
 \dot{\sigma}_3 &= 4g (a_2\sigma_1-a_1\sigma_2).
\end{align}
We have
\begin{align}
 \frac{\mbox{d}}{\mbox{d}t}  \langle \vec{\sigma}\rangle^2=  \frac{\mbox{d}}{\mbox{d}t}(4\sigma_1^2+4\sigma_2^2+\sigma_3^2) =0.
\end{align}
[For this, note that $\sigma_1=\frac{1}{2}\langle \hat{\sigma}_x\rangle$ and $\sigma_2=\frac{1}{2}\langle \sigma_y\rangle$ due to $\hat{\sigma}_{\pm}=\frac{1}{2}(\hat{\sigma}_x\pm \rmi \hat{\sigma}_y)$.] We denote $\ell^2=4\sigma_1^2+4\sigma_2^2+\sigma_3^2$.

We study the stability of fixed points of the above five equations. There are two fixed points, which we label (I) and (II). First, consider fixed point (I) given by
\begin{align}
  a_1 &= a_2 =\sigma_ 2=0,\\
  \sigma_1 &= -\frac{\mathcal{E}}{g}.
\end{align}
For this set of parameters, the flow equations vanish irrespective of the value of $\sigma_3$. We compute the stability matrix
\begin{align}
  M_{ij} = M_{ij}(\vec{c}) =\frac{\partial \dot{c}_i}{\partial c_j}.
\end{align}
At a given fixed point $\vec{c}_\star$, each positive eigenvalue of $M(\vec{c}_\star)$ corresponds to a repulsive direction. A completely stable fixed point only has negative eigenvalues of $M(\vec{c}_\star)$. At fixed point (I), we have
\begin{align}
 M = \begin{pmatrix} -\kappa &0 & 0 & g & 0 \\ 0 & -\kappa & -g  & 0 & 0 \\ 0 & -g\sigma_3 & 0 & 0 & 0 \\ g\sigma_3 & 0 & 0 & 0 & 0 \\ 0 & -4\mathcal{E} & 0 & 0 & 0 \end{pmatrix},
\end{align} 
with eigenvalues
\begin{align}
 &0,\ \frac{1}{2}\Bigl(-\kappa+\sqrt{\kappa^2+4g^2\sigma_3}\Bigr),\ \frac{1}{2}\Bigl(-\kappa-\sqrt{\kappa^2+4g^2\sigma_3}\Bigr).
\end{align}
The latter two eigenvalues are each doubly degenerate. The eigenvalue 0, corresponding to the column which is identically 0, is related to the conservation of $\ell^2$ and is thus unimportant. In the limit $\kappa\to 0$, the sign of $\sigma_3$ matters. For $\sigma_3>0$, we find two positive eigenvalues, and so two repulsive directions. For $\sigma_3<0$, on the other hand, the eigenvalues become purely imaginary with a tiny (negative) real part due to $\kappa$. This describes a stable oscillatory behavior of the expectation values, which are eventually attracted to the fixed point.

Next consider fixed point (II) given by
\begin{align}
 a_1 &= \frac{g}{\kappa}\sigma_2,\\
 a_2 &= -\frac{\mathcal{E}+g\sigma_1}{\kappa},\\
 \sigma_2 &= \pm \sqrt{-\frac{\mathcal{E}}{g}\sigma_1-\sigma_1^2},\\
 \sigma_3 &= 0.
\end{align}
Here the fixed point condition is satisfied irrespective of the value of $\sigma_1$. The eigenvalues of the stability matrix, which are independent of the sign of $\sigma_2$, are given by
\begin{align}
 0,\ -\kappa,\ -\kappa,\ +\frac{2\rmi g}{\kappa}\sqrt{\mathcal{E}(\mathcal{E}+g\sigma_1)},\ -\frac{2\rmi g}{\kappa}\sqrt{\mathcal{E}(\mathcal{E}+g\sigma_1)}.
\end{align}
The eigenvalues remain imaginary (leading to oscillatory behavior) as long as $\mathcal{E} + g \sigma_1>0$. This is solved by either $\sigma_1>0$ or
\begin{align}
 |\sigma_1| < \frac{\mathcal{E}}{g}=\frac{\epsilon}{2}\ \text{for }\sigma_1<0.
\end{align}
Reality of $\sigma_2$, however, requires the second choice with $\sigma_1<0$. Indeed, the mean-field solution shown in Eqs. \eqref{ordermf} satisfies
\begin{align}
    \sigma_1 = -\frac{\ell^2}{2\epsilon} = -\frac{\ell^2}{\epsilon^2} \frac{\epsilon}{2} 
\end{align}
with negative $\sigma_1$ such that $|\sigma_1|<\frac{\epsilon}{2}$ due to $\epsilon^2>\ell^2$ in the symmetry-broken phase.

\vfill

\bibliography{references}

\end{document}